\documentclass[ reprint,
groupedaddress,
 amsmath,amssymb,
 aps,
pra,
onecolumn]{revtex4-2}
\usepackage{graphicx}
\usepackage{comment}
\usepackage{epsfig}
\usepackage{epstopdf}
\usepackage{subcaption}
\usepackage{color}
\usepackage{gensymb}
\usepackage[font=footnotesize,labelformat=simple]{subcaption}

\usepackage{dcolumn}
\usepackage{bm}
\usepackage{amsmath,amssymb}
\usepackage[colorinlistoftodos]{todonotes}

\def\.{\cdot}

\def\_#1{{\bf #1\mit}}
\def\=#1{\overline{\overline #1}}

\graphicspath{{figures/}}
\begin{document}

\title{Temporal Twistronics}

\author{G.~Ptitcyn$^1$}
\email{ptitcyn@seas.upenn.edu}
\author{N.~Engheta$^1$}
\email{engheta@seas.upenn.edu}
%\\
\affiliation{$^1$University of Pennsylvania, Department of Electrical and Systems Engineering, Philadelphia, PA 19104, U.S.A.}

\begin{abstract}
The concept of twistronics and moir\'e physics, which is present in twisted two-dimensional bilayer materials, has recently attracted growing attention in various fields of science and engineering such as condensed matter physics, nanophotonics, polaritonics and excitonics.  The twist angle between the two layers has offered an additional degree of control over electron and photon interaction with such structures.  Inspired by the photonic version of twistronics, here we introduce and investigate theoretically the temporal analogue of twistronics in anisotropic optical media.   We study how a monochromatic electromagnetic plane wave propagating in a spatially unbounded, anisotropic medium undergoes major changes when the relative permittivity tensor of the medium is rapidly changed in time to create a new anisotropic medium that is the rotated version of the original medium.  We consider both the elliptic and hyperbolic anisotropic scenarios.  The propagation-angle-dependent forward (FW) and backward (BW) waves with their converted frequencies and relative amplitudes are obtained.  To concentrate on the main features of this concept without getting into details of dispersion, in our work here we assume dispersionless and lossless material parameters.  Our results reveal how frequency conversion is highly dependent on the direction of propagation of the original wave, rotation angle, and initial values of the material parameters, proposing another class of "magic angles" for such temporal twistronics.      

% reduced abstract
%The concept of twistronics and moir\'e physics, which is present in twisted two-dimensional bilayer materials, has recently attracted growing attention in various fields of science and engineering such as condensed matter physics, superconductivity, nanophotonics, polaritonics and excitonics.   Inspired by the photonic version of twistronics, here we introduce and investigate theoretically the temporal analogue of twistronics in anisotropic optical media.   We study how a monochromatic electromagnetic plane wave propagating in a spatially unbounded, anisotropic medium undergoes major changes when the relative permittivity tensor of the medium is rapidly changed in time to create a new anisotropic medium that is the rotated version of the original medium.  We consider both the elliptic and hyperbolic anisotropic scenarios.  Our results reveal how frequency conversion is highly dependent on the direction of propagation of the original wave, proposing another class of "magic angle" for such temporal twistronics.      

\end{abstract}

	\maketitle
% \section{Introduction}

Study of light-matter interaction in anisotropic crystals has a long history~\cite{fresnel1821note,yariv1983optical}.  In addition to the elliptic anisotropy common in conventional uniaxial and biaxial crystals, in recent years the notion of hyperbolic materials in the context of metamaterials and metasurfaces has been explored extensively~\cite{poddubny2013hyperbolic,ferrari2015hyperbolic,shekhar2014hyperbolic,gomez2016flatland,high2015visible}.  These structures, which can be formed by periodic arrangements of layers of dielectric with $\epsilon_{\mathrm{layer}~1}>0$ and metal with $\epsilon_{\mathrm{layer}~2}<0$, have received considerable attention due to their salient electromagnetic characteristics.  In particular, due to the hyperbolic nature of isofrequency curve in their dispersion diagrams, such media exhibit extreme anisotropy, for example, having high values of wave numbers along the asymptotes of the dispersion hyperbolae~\cite{huo2019hyperbolic,guo2020hyperbolic}.

In an entirely different paradigm, the notion of twisted two-dimensional material bilayer has been investigated extensively in recent years, demonstrating exciting features via control of the twist angles between the two layers~\cite{carr2017twistronics}.  Such twistronics and moir\'e physics have been studied for electronic transport~\cite{ren2020twistronics}, photonics~\cite{hu2021twistronics,hu2020topological,tang2023experimental}, polaritonics~\cite{duan2020twisted,hu2020topological}, and excitonics~\cite{ciarrocchi2022excitonic}, to name a few. 

Another path for manipulating and sculpting light-matter interaction in metamaterials and metasurfaces is the notion of time-varying media, i.e., four-dimensional (4D) metamaterials, in which material parameters (e.g., relative permittivity) can be changed rapidly in time, in addition to (or instead of) their spatial variation in space, while the wave is in the material.  Such spatiotemporal modulation of material media, which has a long history dating back to 1950s~\cite{zadeh1950frequency,zadeh1950determination,morgenthaler1958velocity,cullen1958travelling,tien1958parametric}, has recently received renewed attention and growing interest in many groups worldwide, offering exciting possibilities for wave manipulation with various potential applications~\cite{engheta2020metamaterials,engheta2023four,tirole2023double,galiffi_photonics_2022,zhou2020broadband,pacheco2020temporal,quinones_tunable_2021,yin2022efficient,biancalana_dynamics_2007,zurita-sanchez_reflection_2009,reyes-ayona_observation_2015,lustig_topological_2018,park_spatiotemporal_2021,sharabi2021disordered,GregAtom,boltasseva2024photonic}.

In the present work, we merge the above three concepts by introducing the notion of ${\emph{temporal twistronics}}$ or ${\emph{temporal moir\'e}}$ in lossless, dispersionless anisotropic media. We theoretically study how a monochromatic plane wave in an anisotropic medium with real-valued permittivity tensor is affected by the rapid change of the material parameters that creates a new anisotropic medium as the replica of the original medium but with its optical axes rotated by a certain temporal twist angle. We consider the elliptic and hyperbolic anisotropy for such temporal twistronics.  In the work presented here, we assume no frequency dispersion for the material parameters.

 \begin{figure*}
\centering
\begin{minipage}{0.39\textwidth}
\begin{subfigure}{1\textwidth}
\includegraphics[width=1\textwidth]{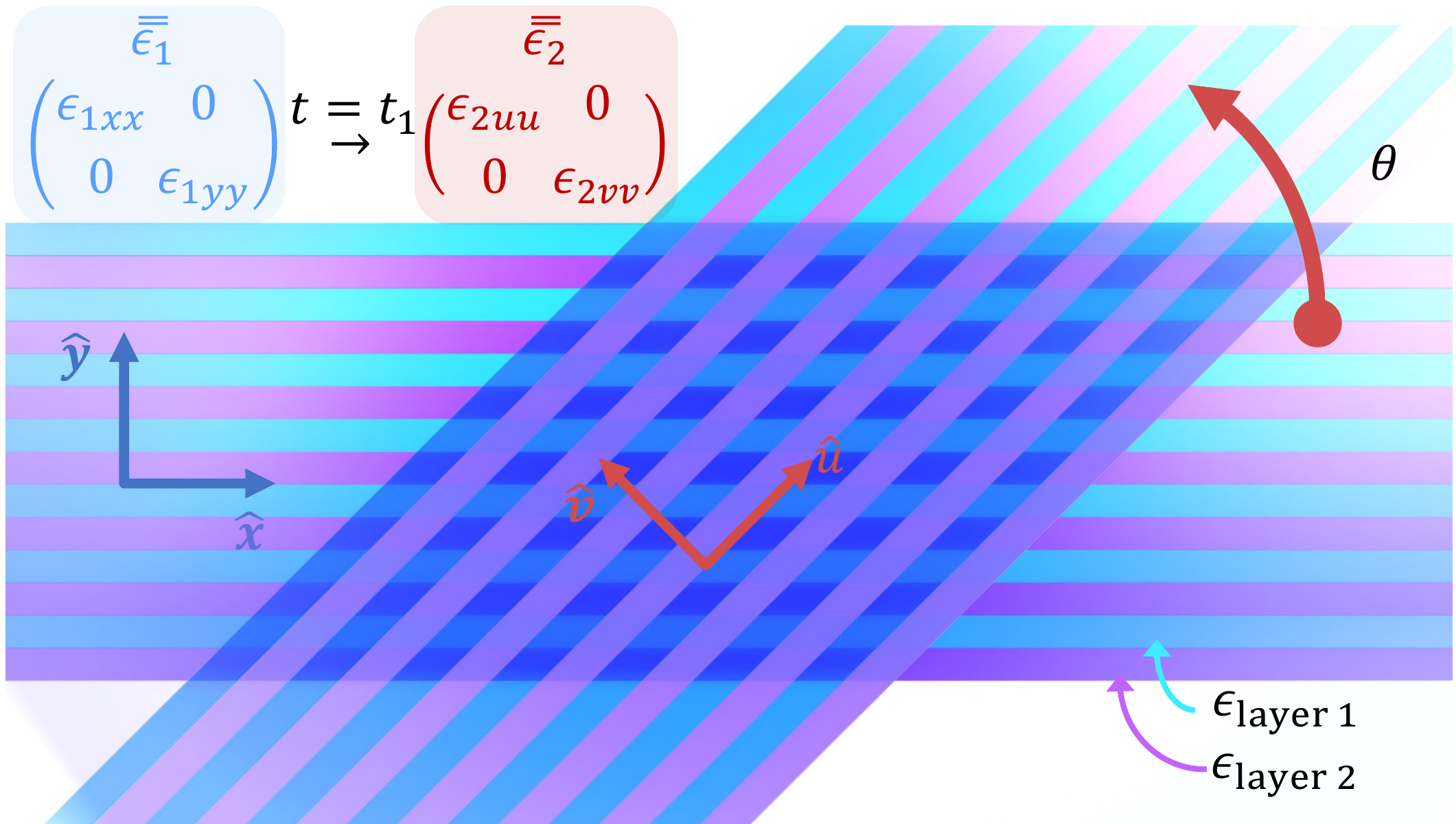}
\caption{}
\end{subfigure}
\end{minipage}
\begin{minipage}{0.59\textwidth}
\begin{subfigure}{0.45\textwidth}
\centering
\includegraphics[width=\textwidth]{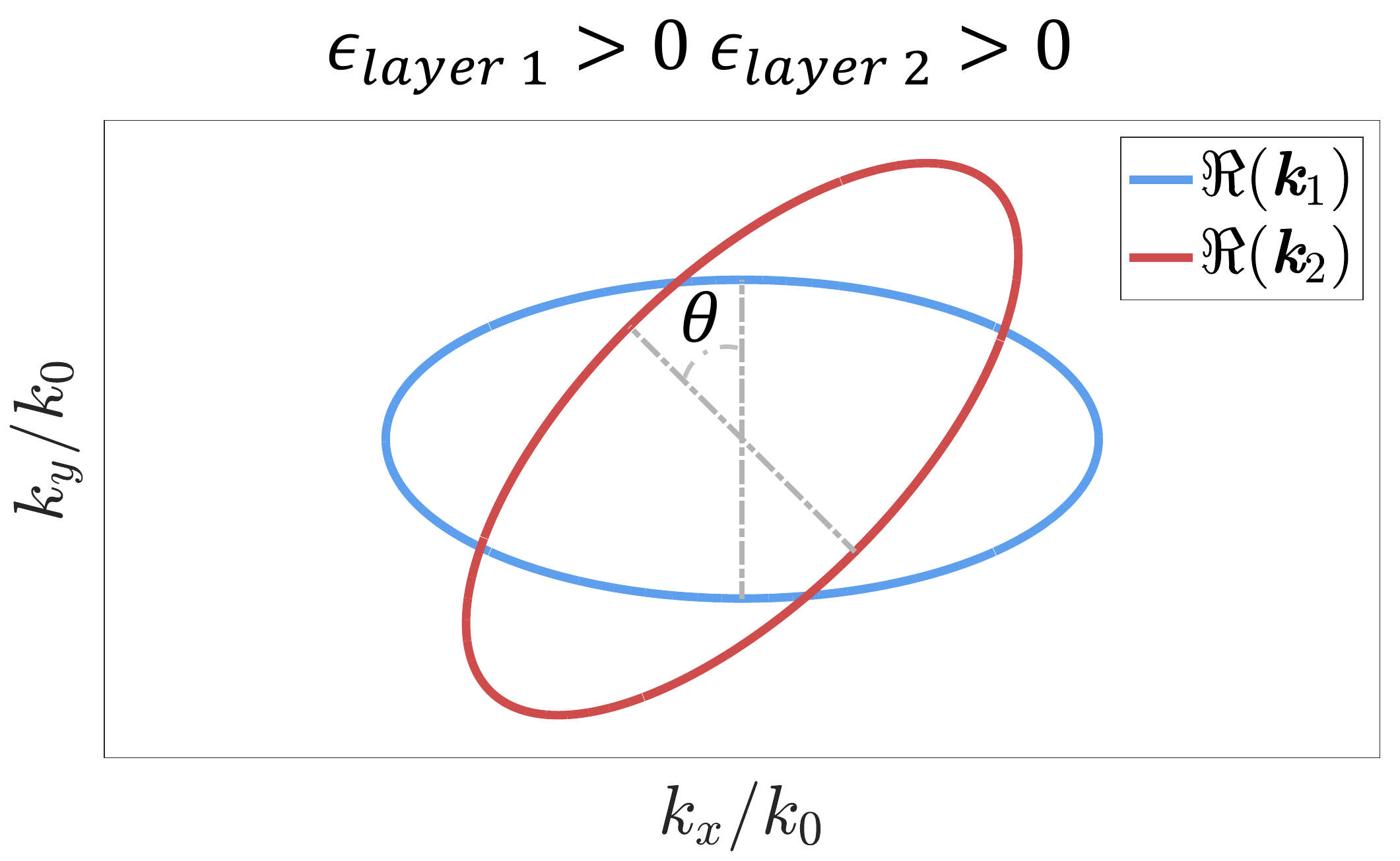}
\caption{}
\end{subfigure}
~
\begin{subfigure}{0.45\textwidth}
\centering
\includegraphics[width=\textwidth]{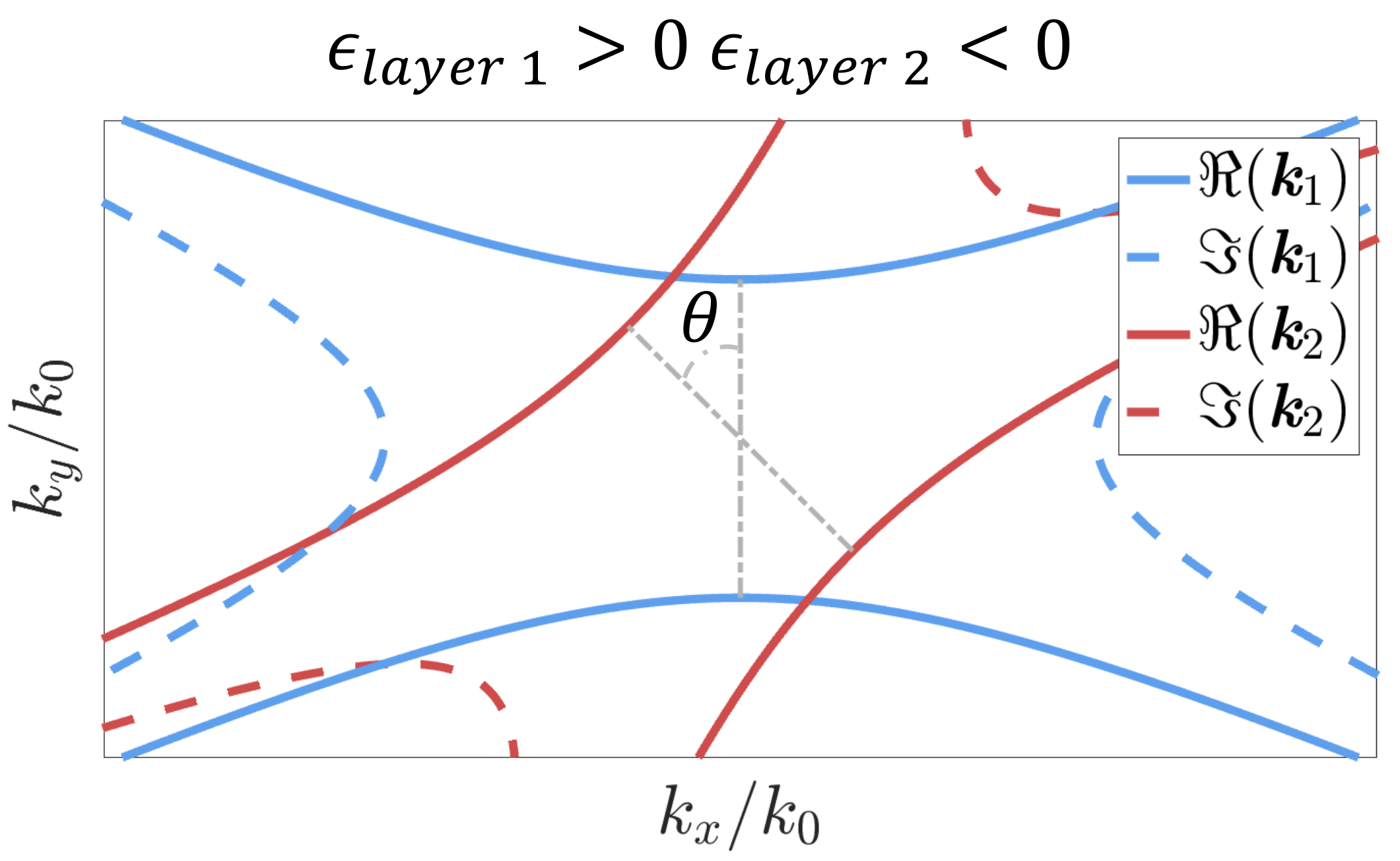}
\caption{}
\end{subfigure}\\
\begin{subfigure}{0.45\textwidth}
\centering
\includegraphics[width=\textwidth]{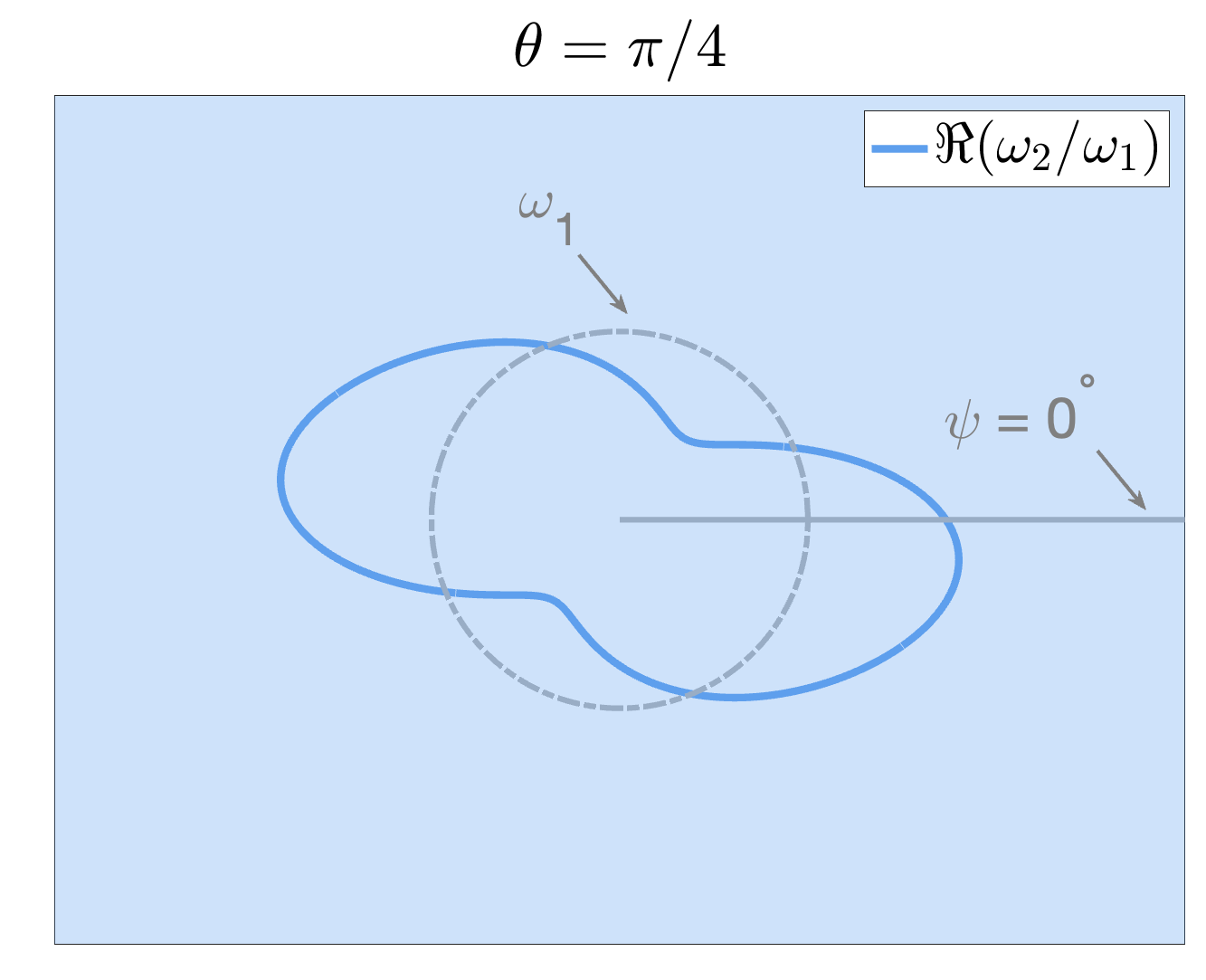}
\caption{}
\end{subfigure}
~
\begin{subfigure}{0.45\textwidth}\includegraphics[width=\textwidth]{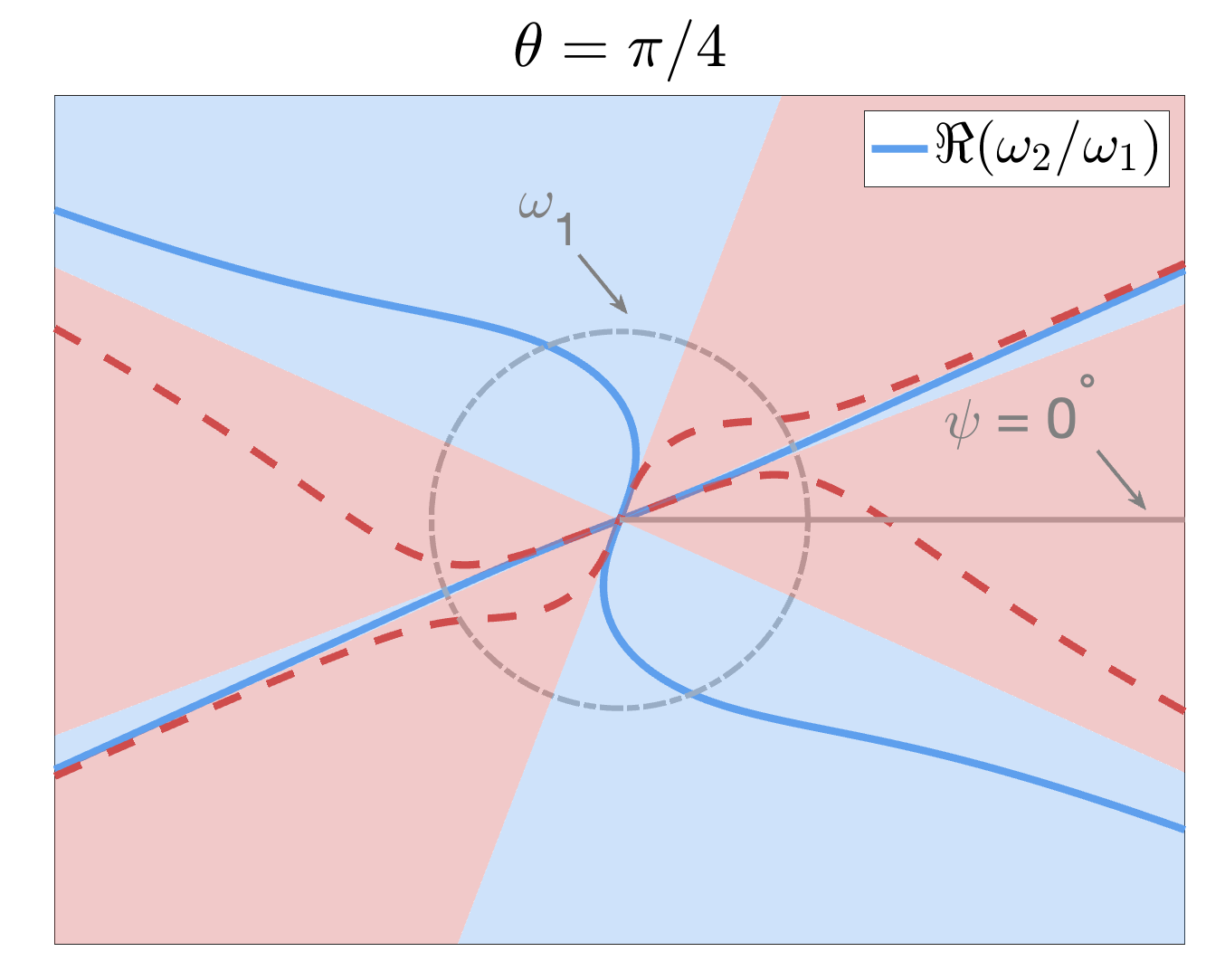}
\caption{ }
\end{subfigure}
\end{minipage}    
\caption{(a) Conceptual representation of anisotropic medium with its rotated version at angle $\theta$. (b)~and (c) Isofrequency plots in polar coordinates for elliptic medium ($\epsilon_{\mathrm{layer}~1}>0$ and $\epsilon_{\mathrm{layer}~2}>0$) and hyperbolic medium ($\epsilon_{\mathrm{layer}~1}>0$ and $\epsilon_{\mathrm{layer}~2}<0$), respectively. (d) and (e) Normalized converted frequency of the wave as a function of direction of propagation, $\psi$ (plotted in polar coordinates), after an abrupt temporal rotation-like change of material parameters (here shown for rotation $\theta$=$\pi$/4) for (d) elliptic medium ($\epsilon_{1xx}=1$, $\epsilon_{1yy}=5$) and (e) hyperbolic medium ($\epsilon_{1xx}=1$, $\epsilon_{1yy}=-5$). Gray circle shows the original frequency before the jump. Blue and red areas, respectively, show regions where $\omega_2$ is real and purely imaginary. The polar angle in these coordinates represent the direction of propagation, $\psi$, of the original plane wave.}
\label{fig_concept}
\end{figure*}

% and isofrequency curves corresponding to elliptic ($\epsilon_{\mathrm{layer}~1}>0$ and $\epsilon_{\mathrm{layer}~2}>0$) and hyperbolic ($\epsilon_{\mathrm{layer}~1}>0$ and $\epsilon_{\mathrm{layer}~2}<0$) cases

%\section{Formulation of the Problem}
Consider a monochromatic transverse-magnetic (TM) electromagnetic plane waves
%with its magnetic field vector parallel with the $z$ axis and 
with angunar frequnecy $\omega_1$ and its wave vector $\boldsymbol{k}$ in the $x$-$y$ plane (making angle $\psi$ with respect to the $x$ axis)
\begin{subequations}
 \begin{align}
    \_H_1&=\_{\hat{z}} e^{i(\omega_1 t-k_x x-k_y y)},\\
    \_E_1&=\frac{1}{\epsilon_0}\bigg(\frac{k_y\_{\hat{x}}}{-\epsilon_{1xx}\omega_1}+\frac{k_x\_{\hat{y}}}{\epsilon_{1yy}\omega_1}\bigg)e^{i(\omega_1 t-k_x x-k_y y)},
\end{align}
\label{eq_fields_1}
\end{subequations}
propagating in an unbounded homogeneous lossless anisotropic medium (assuming its optical axes along the $x$-$y$ coordinates) with the initial relative permittivity tensor 
\begin{align}
    \={\epsilon}_1=\Bigg(\begin{array}{cc}
        \epsilon_{1xx} & 0 \\
         0 & \epsilon_{1yy}
    \end{array}\Bigg).
    \label{eq_anisotropic_permittivity_tensor}
\end{align} 
Vacuum permittivity is denoted as $\epsilon_0$. Since the electric field is perpendicular to $\hat{\mathbf{z}}$, the permittivity tensor is simply shown as a $2 \times 2$ matrix.  Moreover, we assume that the medium under study is nonmagnetic with permeability $\mu_0$. 

\begin{comment}
It is well known that rotating this medium with angle $\theta$ in counterclockwise direction leads to a non-diagonal permittivity tensor $\epsilon_{2}$  
\begin{align}
    \={\epsilon}_2=Q\cdot\={\epsilon}_1\cdot Q^{-1}=\Bigg(\begin{array}{cc}
        \epsilon_{2xx} & \epsilon_{2xy} \\
         \epsilon_{2yx} & \epsilon_{2yy}
    \end{array}\Bigg),
    \label{eq_permittivity2}
\end{align}
where $Q$ is the unitary rotation matrix given as
\begin{align}
    Q=\Bigg(\begin{array}{cc}
        \cos{\theta} & -\sin{\theta} \\
         \sin{\theta} & \cos{\theta}
    \end{array}\Bigg).
\end{align}
The elements of the rotated tensor in \eqref{eq_permittivity2} can be expressed as
\begin{subequations}
\begin{align}
    \epsilon_{2xx}&=\frac{1}{2}(\epsilon_{1xx}+\epsilon_{1yy}+(\epsilon_{1xx}-\epsilon_{1yy})\cos{2\theta}),\\
    \epsilon_{2yy}&=\frac{1}{2}(\epsilon_{1xx}+\epsilon_{1yy}+(\epsilon_{1yy}-\epsilon_{1xx})\cos{2\theta}),\\
    \epsilon_{2xy}&=\epsilon_{2yx}=\frac{1}{2}(\epsilon_{1xx}-\epsilon_{1yy})\sin{2\theta}.
\end{align}
\end{subequations}
\end{comment}

Let us analytically study the case shown in Fig.~\ref{fig_concept}, i.e. abrupt change of material parameters in time such that the material after this temporal change resembles the original anisotropic medium, but with its optics axes rotated with respect to the original configuration. Throughout the paper we refer to it as temporal rotation-like material change in order to stress that the medium itself remains motionless and only material properties abruptly change in time, mimicking an instantaneous switch-like rotation. We focus our attention on the temporal step change of material properties, which allow us to consider it as a temporal boundary.  In our work here we consider two types of lossless anisotropic materials: elliptic medium (both $\epsilon_{1xx}$ and $\epsilon_{1yy}$ positive); and hyperbolic medium (one of $\epsilon_{1xx}$ and $\epsilon_{1yy}$ positive and the other one negative) (see Fig.~\ref{fig_concept}). In the rotated frame with the new basis vectors $\hat{\mathbf{u}}$ and $\hat{\mathbf{v}}$ the permittivity tensor $\={\epsilon}_2$ remains diagonal with permittivities $\epsilon_{2uu}=\epsilon_{1xx}$ and $\epsilon_{2vv}=\epsilon_{1yy}$.  The new basis vectors can be expressed in terms of the old basis vectors $\mathbf{\hat{x}}$ and $\mathbf{\hat{y}}$, as
\begin{align}
\Bigg(\begin{array}{c}
         \hat{\mathbf{u}}\\
         \hat{\mathbf{v}}
    \end{array}\Bigg)=Q^T\cdot
    \Bigg(\begin{array}{c}
         \hat{\mathbf{x}}\\
         \hat{\mathbf{y}}
    \end{array}\Bigg),
\label{eq_u_v_through_Q_x_y}
\end{align}
where $Q$ is the unitary rotation matrix given as
\begin{align}
    Q=\Bigg(\begin{array}{cc}
        \cos{\theta} & -\sin{\theta} \\
         \sin{\theta} & \cos{\theta}
    \end{array}\Bigg).
\end{align}

It is worth mentioning that the wave vector $\textbf{\textit{k}}$ must be preserved before and after the temporal step change, i.e., $\textbf{\textit{k}}_1=\textbf{\textit{k}}_2 \equiv \textbf{\textit{k}}$, while the frequency changes. In the initial medium the wave vector is specified by its $x$ and $y$ components, $k_{x}$ and $k_{y}$, for the wave with the original frequency  $\omega_1$. Figures~\ref{fig_concept}(b) and~(c) show isofrequency curves for elliptic case with 
$\epsilon_{1xx}=1$ and  $\epsilon_{1yy}=5$ and for hyperbolic cases with $\epsilon_{1xx}=1$ and  $\epsilon_{1yy}=-5$. Therefore we fix the frequency of the initial wave and find the corresponding wave vector. In the rotated frame the wave vector can be expressed as
\begin{align}
    \Bigg(\begin{array}{c}
         k_{u}\\
         k_{v}
    \end{array}\Bigg)=Q^T\cdot
    \Bigg(\begin{array}{c}
         k_x\\
         k_y
    \end{array}\Bigg).
    \label{eq_kukv_through_kxky}
\end{align}
Using notation of the new (rotated) frame, the frequency after the temporal jump, denoted by $\omega_2$, can be found as
\begin{align}
    \omega_2=\pm c \sqrt{\frac{k_u^2}{\epsilon_{2vv}}+\frac{k_v^2}{\epsilon_{2uu}}},
\end{align}
\begin{comment}
\begin{align}
\Bigg|
\begin{array}{cc}
    \frac{\omega_2^2}{c^2}\epsilon_{2xx}-k_y^2 & \frac{\omega_2^2}{c^2}\epsilon_{2xy} + k_xk_y \\
    \frac{\omega_2^2}{c^2}\epsilon_{2xy} + k_xk_y & \frac{\omega_2^2}{c^2}\epsilon_{2yy}-k_x^2
\end{array}
\Bigg|=0.
\end{align}
\end{comment}

%\begin{figure}
%\centering
%\begin{subfigure}{0.23\textwidth}
%\centering
% \includegraphics[width=\textwidth]{wave_vector_elliptical.png}
%\caption{}
%\end{subfigure}
%~~
%\begin{subfigure}{0.23\textwidth}
%\centering
% \includegraphics[width=\textwidth]{wave_vector_hyperbolic.png}
%\caption{}
%\end{subfigure}\\
%\begin{subfigure}{0.23\textwidth}
%\centering
% \includegraphics[width=\textwidth]{Frequency_w2_elliptical.png}
%\caption{}
%\end{subfigure}
~~

\begin{figure*}
\centering
\begin{subfigure}{0.32\textwidth}
\centering
 \includegraphics[width=\textwidth]{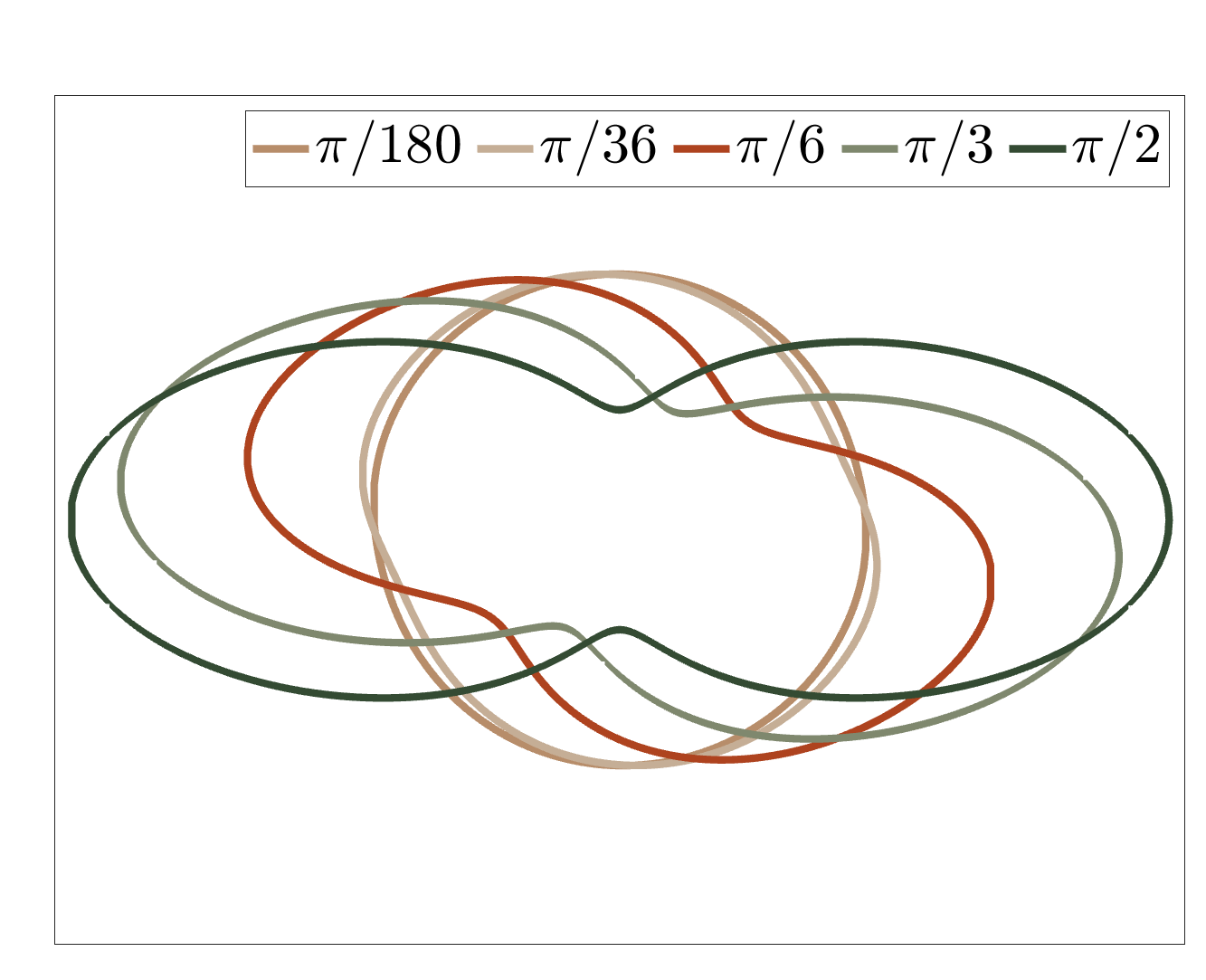}
\caption{}
\end{subfigure}
~
\begin{subfigure}{0.32\textwidth}
\centering
 \includegraphics[width=\textwidth]{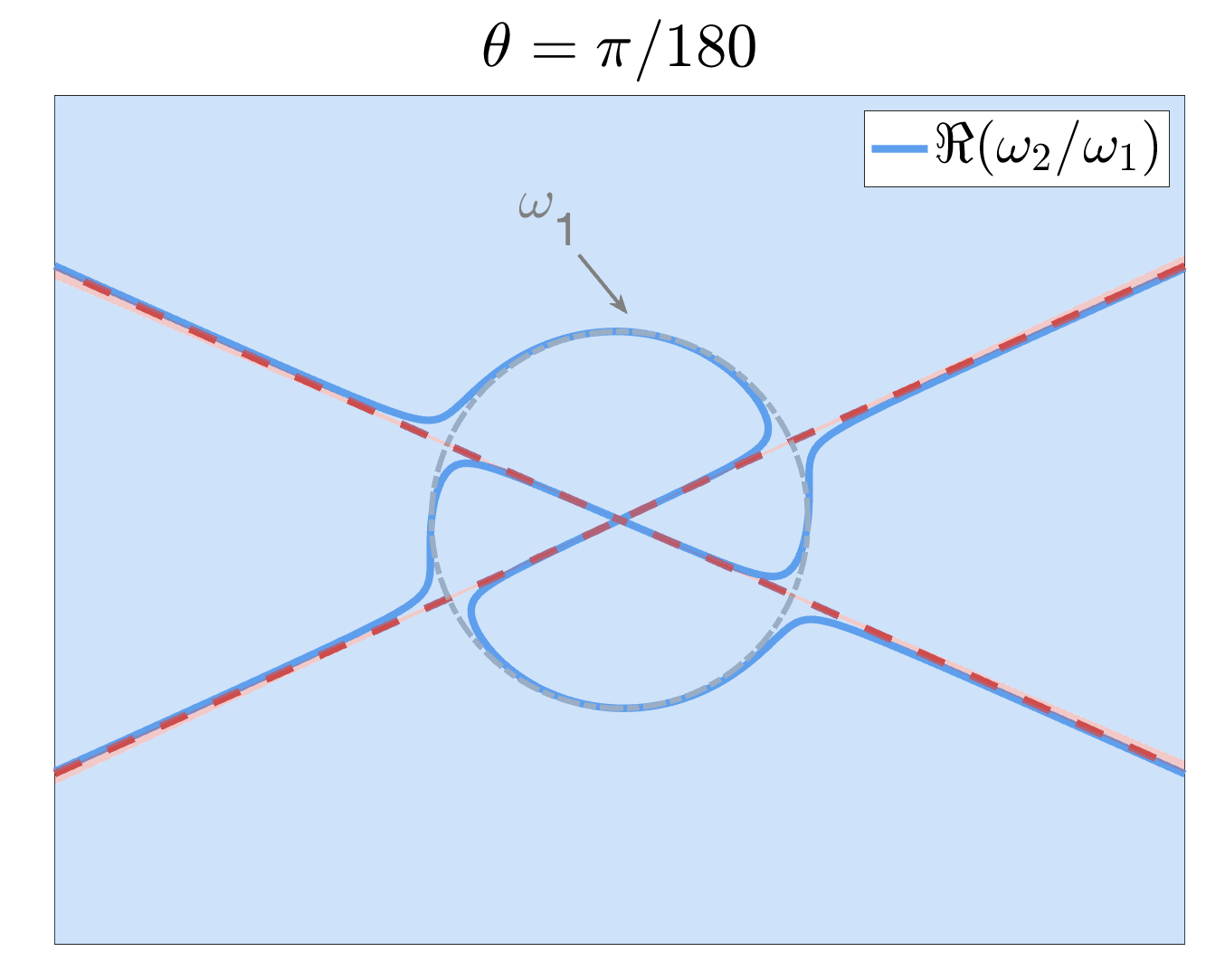}
\caption{}
\end{subfigure}
~
\begin{subfigure}{0.32\textwidth}
\centering
 \includegraphics[width=\textwidth]{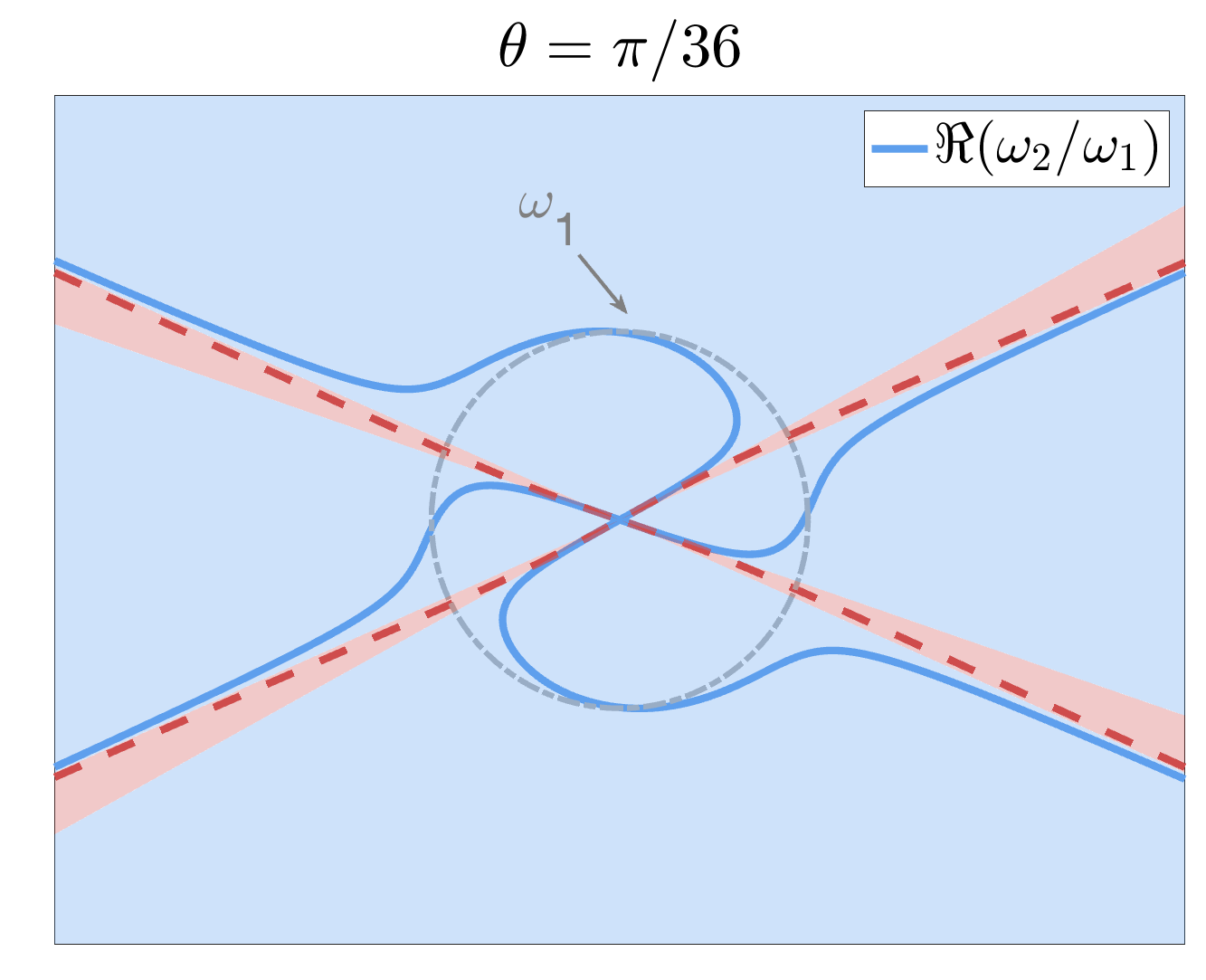}
\caption{}
\end{subfigure}\\
\begin{subfigure}{0.32\textwidth}
\centering
 \includegraphics[width=\textwidth]{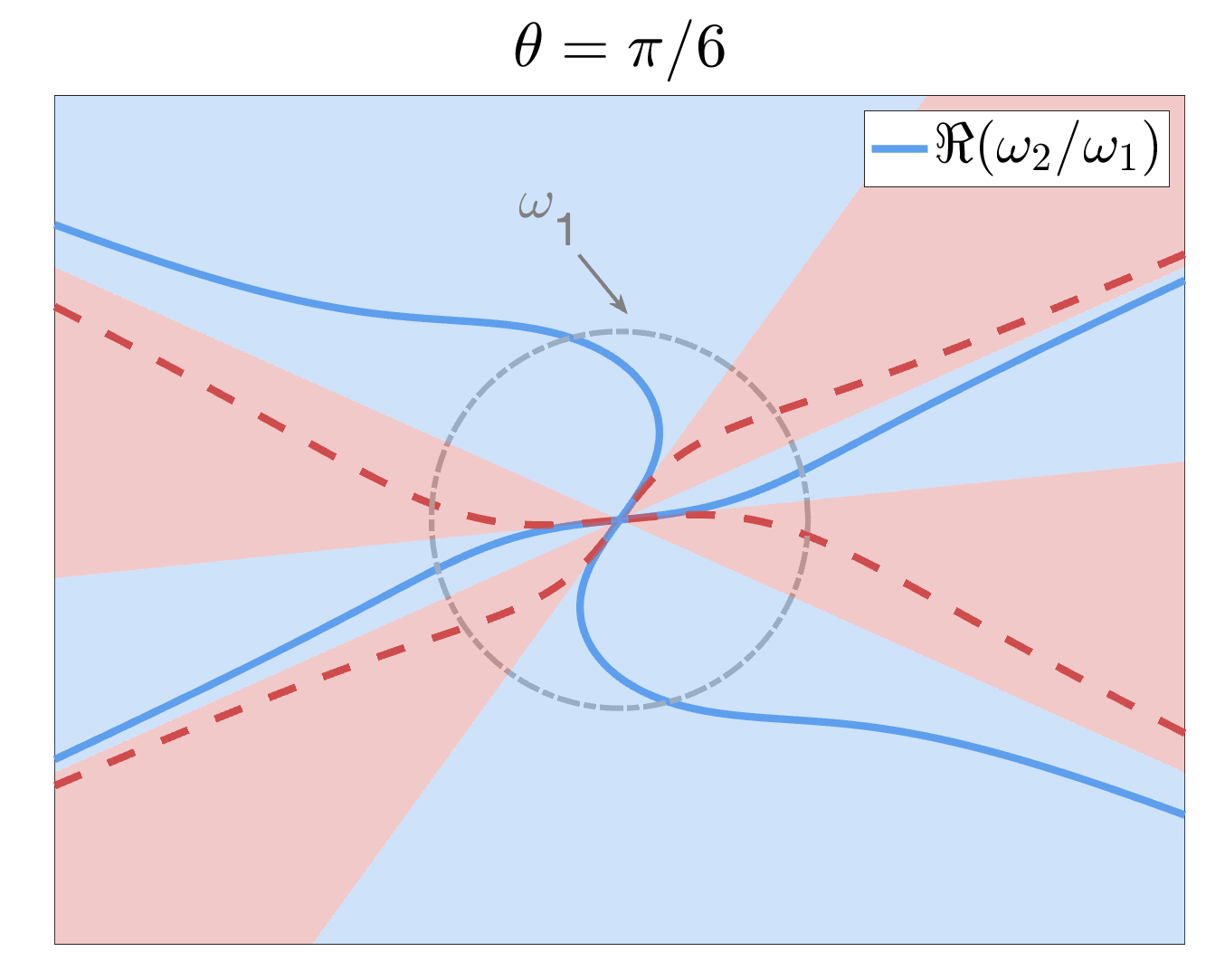}
\caption{}
\end{subfigure}
~
\begin{subfigure}{0.32\textwidth}
\centering
 \includegraphics[width=\textwidth]{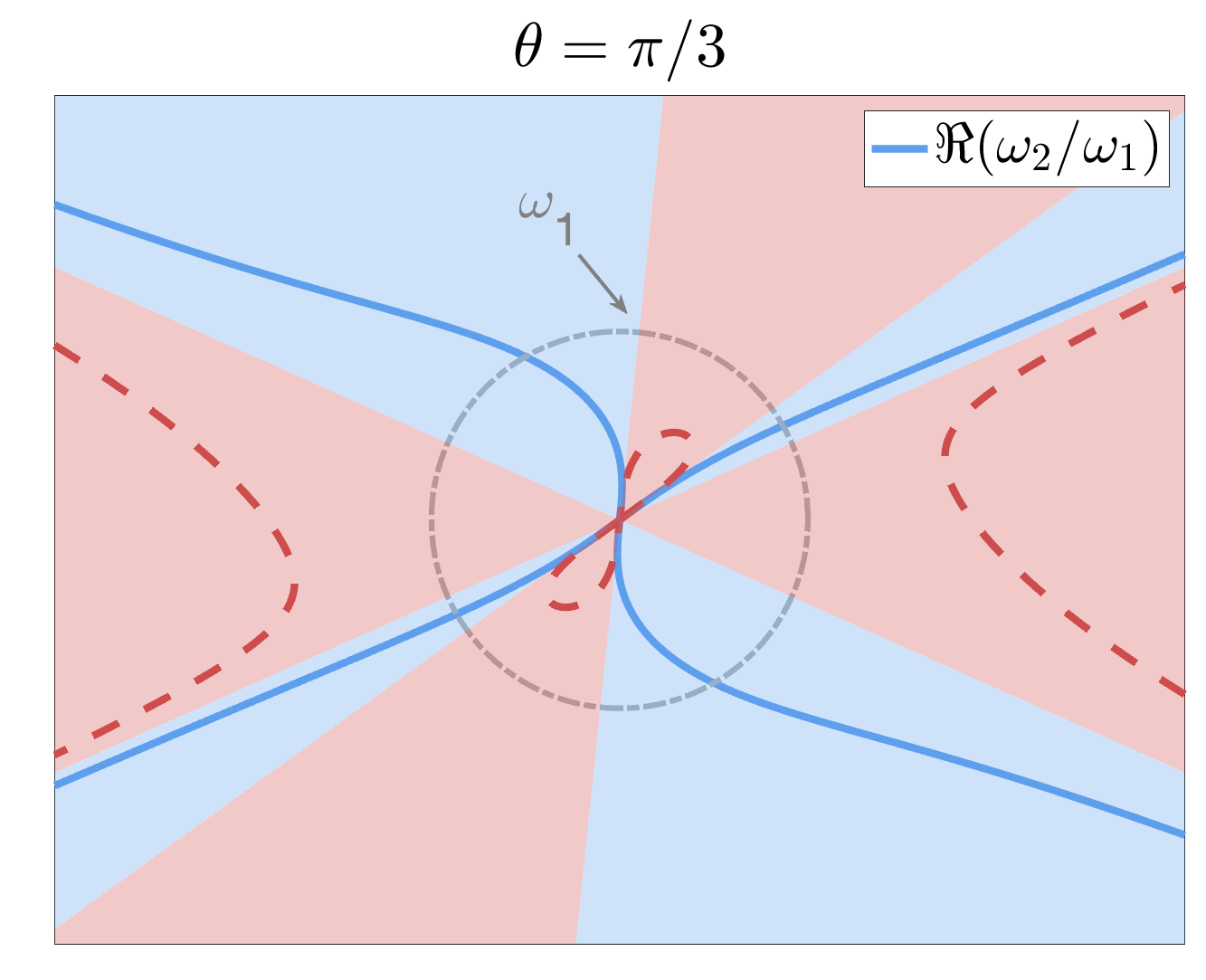}
\caption{}
\end{subfigure}
~
\begin{subfigure}{0.32\textwidth}
\centering
 \includegraphics[width=\textwidth]{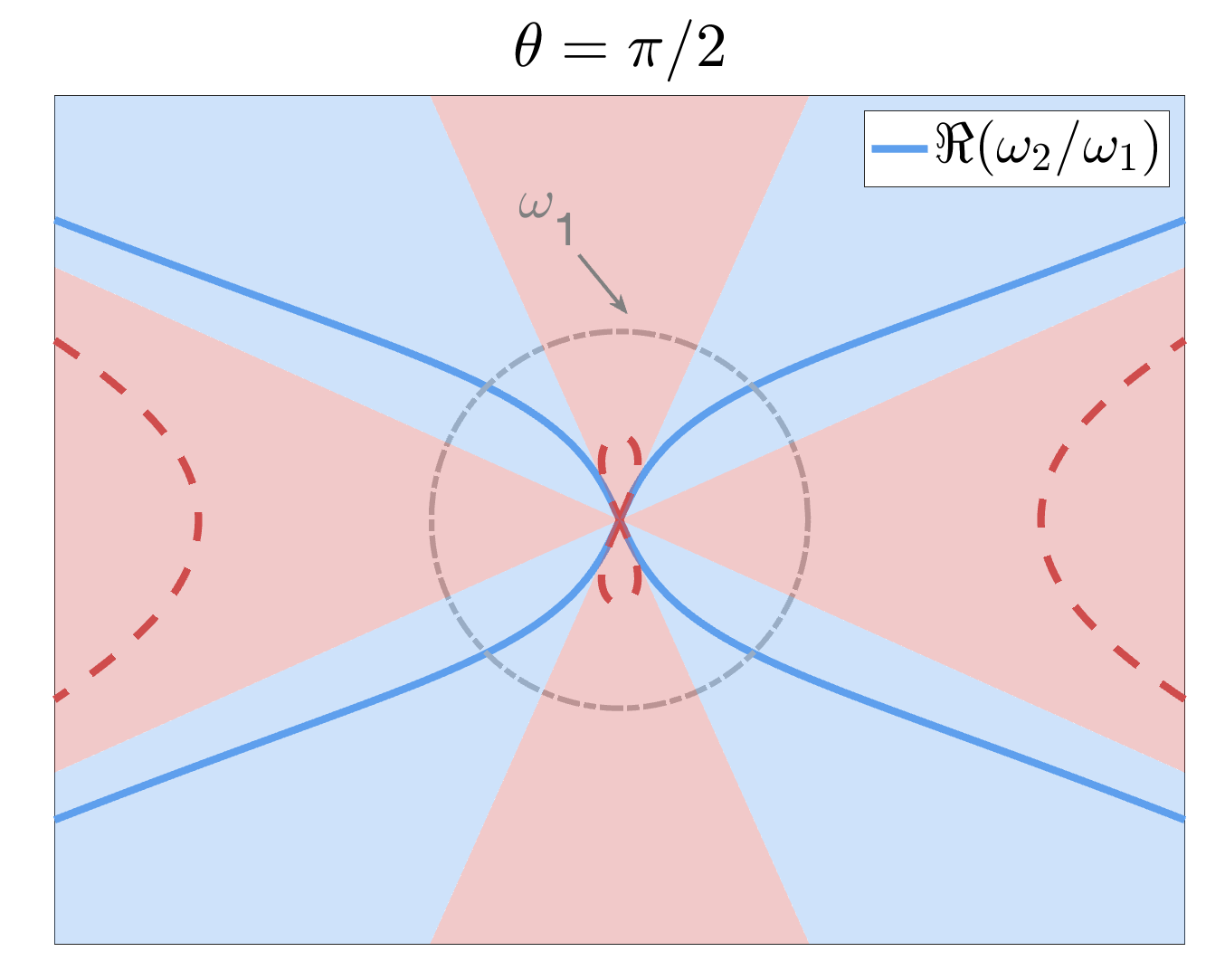}
\caption{}
\end{subfigure}
\\
\begin{subfigure}{0.32\textwidth}
\centering
 \includegraphics[width=\textwidth]{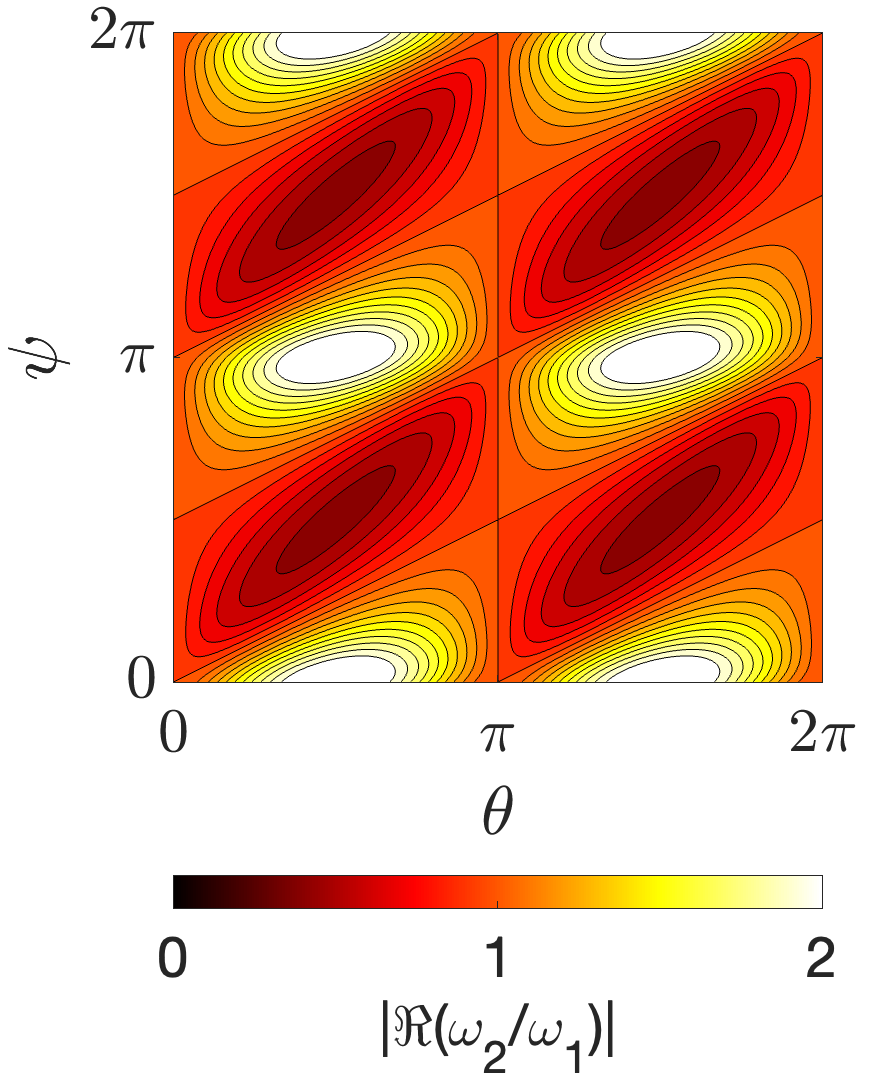}
\caption{}
\end{subfigure}
~
\begin{subfigure}{0.32\textwidth}
\centering
 \includegraphics[width=\textwidth]{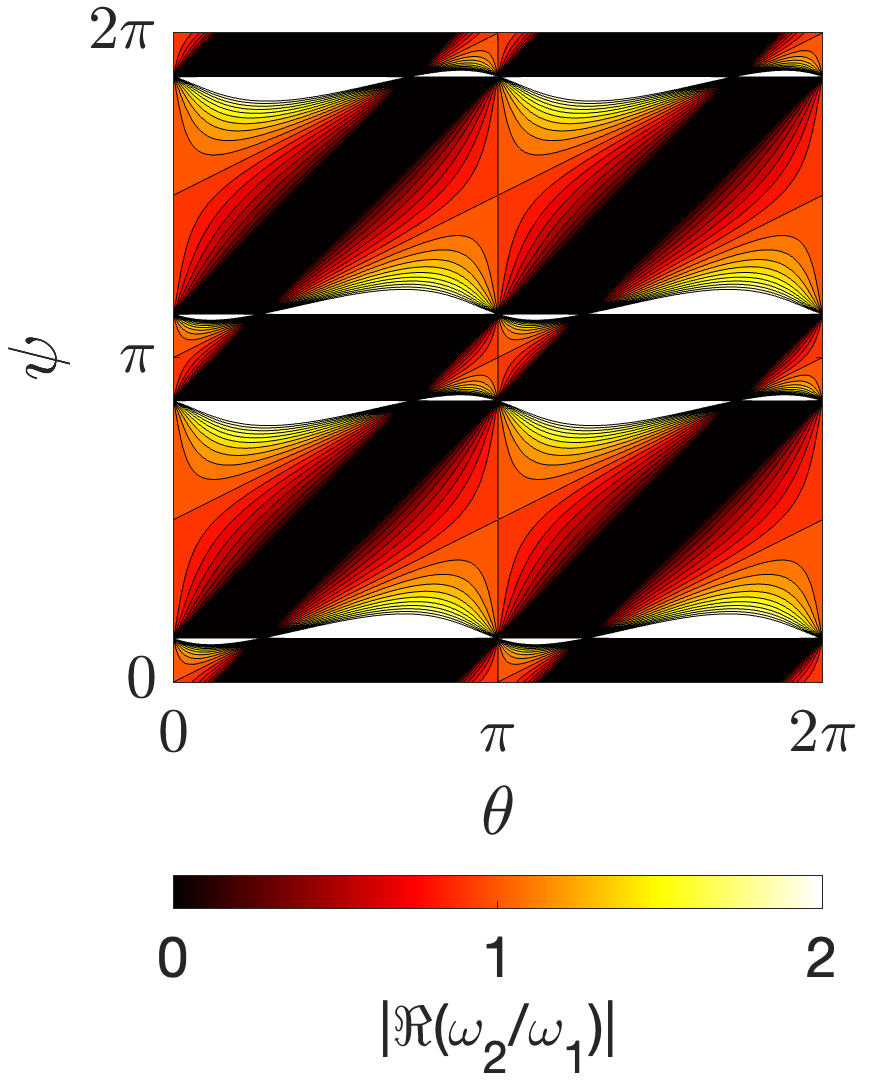}
\caption{}
\end{subfigure}
~
\begin{subfigure}{0.32\textwidth}
\centering
 \includegraphics[width=\textwidth]{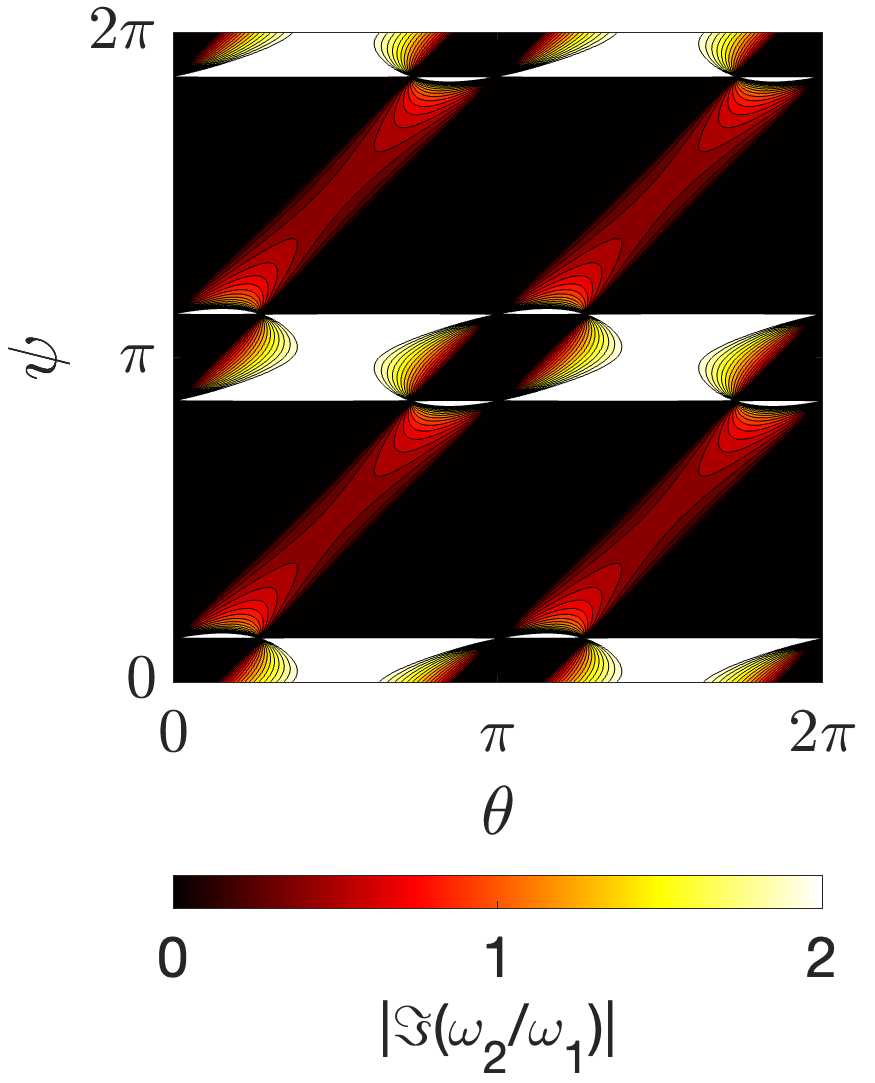}
\caption{}
\end{subfigure}
\caption{(a-f) Polar plots of the converted frequencies $\omega_2$ (normalized to $\omega_1$) as a function of the direction of propagation, $\psi$, after the temporal jump for (a) elliptic case ($\epsilon_{1xx}=1$, $\epsilon_{1yy}=5$) and (b-f) hyperbolic case ($\epsilon_{1xx}=1$, $\epsilon_{1yy}=-5$) for rotation angles, $\theta$, between 1$\degree$ and 90$\degree$, which are shown in each figure. Blue and red areas in (b-f), respectively, show where $w_2$ is purely real and purely imaginary. (g-i) Contour plots of $\omega_2$ (normalized to $\omega_1$) as a function $\psi$, and rotation angle $\theta$ for (g) elliptic case and (h,i) hyperbolic case with real part in (h) and imaginary part in (i).}
\label{fig_frequency_omega2_many}
\end{figure*}

%It is worth noting that as expected this expression also provides the well known result for the case with diagonal permittivity tensor (i.e., the off-diagonal terms being zero, $\epsilon_{2xy}=0$)~\cite{pacheco2020temporal}.

%\section{Results and Discussion}

Figure~\ref{fig_concept}(d) and~(e) show the converted frequency $\omega_2$ (normalized with respect to $\omega_1$) as a function of direction of propagation, $\psi$, in the polar coordinate system for lossless elliptic and hyperbolic cases after temporal rotation-like material change with angle $\theta=\pi/4$. It is interesting to note that in the hyperbolic case for certain directions the converted frequency $\omega_2$ becomes purely imaginary, indicating one of the two following scenarios: (a) when before the temporal change the initial $k-$vector points into allowed angular directions (i.e., $\textit{\textbf{k}}_2$ being real) of the hyperbolic medium  while after the temporal change it  points into forbidden angular directions (i.e., $\textit{\textbf{k}}_2$ being imaginary) of the medium, and (b) the other way around, i.e., when before the temporal change the initial $k-$vector points into the forbidden angular directions, and then after the change it points into the allowed angular directions. Therefore, to ensure preservation of wave vector $k$, frequency $\omega_2$ becomes purely imaginary, causing the signal to grow and decay with time.  This is due to the fact that in our study here we have assumed no dispersion for the material parameters of the hyperbolic medium.  Such dispersionless assumption for the negative permittivity requires use of non-Foster approach~\cite{pacheco2023holding}. We are currently studying the role of dispersion in temporal hyperbolic media, the results of which will be reported in a future publication.

To highlight the role of the rotation angle on the converted frequency, Fig.~\ref{fig_frequency_omega2_many}(a-f) shows a set of polar plots of converted frequency $\omega_2$ (normalized with respect to $\omega_1$) after the temporal jump as a function of the direction of propagation, $\psi$, for several different rotation angles, $\theta$, for the elliptic and hyperbolic scenarios. Figure~\ref{fig_frequency_omega2_many}(g-i) shows normalized converted frequency depending on $\psi$ and rotation angle $\theta$. Although our original wave had a single frequency regardless of its direction of propagation, the converted frequency of the wave after the temporal jump attains values that depend on $\psi$. For elliptic case frequency $\omega_2$ remains always real and it can be varied within a certain range using angles $\theta$ and $\psi$, which is consistent with what was found in Ref.~\cite{akbarzadeh2018inverse}, whereas for hyperbolic case converted $\omega_2$ looks far more complicated. Two interesting features in the hyperbolic case can be highlighted: (1) Under the dispersionless assumption for the real-valued material parameters, the converted frequency in the hyperbolic case can reach extremely high and near-zero values.  This is due to the fact that around the asymptotes in the isofrequency curves the wavenumber has values that can vary a lot with a small change in the direction of propagation.  So when the temporal rotation-like material change happens and the directions of the asymptotes change, in order to preserve the wave vector before and after the jump, the frequency should change accordingly.  This change in the converted frequency can also be a lot with a small change in the asymptote rotation.  This is why we note that the converted frequency may attain values near zero or very high; (2) Again under the assumption of no dispersion for the real-valued permittivity tensors (i.e., the non-Foster assumption), the converted frequency for certain directions of propagation is purely imaginary.  As mentioned before, this is due to the fact that the temporal rotation-like material change can rotate the forbidden angular directions of propagation.  So if the original direction of propagation of our wave with frequency $\omega_1$ is along the forbidden (or allowed) direction and if the temporal change makes this direction to be along the allowed (or forbidden) direction in the rotated hyperbolic material, then the converted frequency will become purely imaginary in order to preserve the continuity of wave vector before and after the temporal jump.

To shed light on the wave phenomena occurring due to an abrupt rotation-like material change in time one needs to determine amplitudes of the waves after the jump. The fields in the rotated medium after the temporal jump at moment $t=t_1$ will consist of forward~(FW) and backward~(BW) propagating waves~\cite{morgenthaler1958velocity}. In the rotated frame these fields can be written as 
\begin{subequations}
\begin{align}
    \_H_2&=\_{\hat{z}}e^{i(-k_u u-k_v v)} (Ae^{i\omega_2(t-t_1)}-Be^{-i\omega_2(t-t_1)}),\\
    \_E_2&=\frac{1}{\epsilon_0}e^{i(-k_u u-k_v v)}(Ae^{i\omega_2(t-t_1)}+Be^{-i\omega_2(t-t_1)})
   \bigg(\frac{k_v\_{\hat{u}}}{-\epsilon_{2uu}\omega_2}+\frac{k_u\_{\hat{v}}}{\epsilon_{2vv}\omega_2}\bigg).
\end{align}
\label{eq_fields_2}
\end{subequations}
Using Eqs.~\eqref{eq_u_v_through_Q_x_y} and~\eqref{eq_kukv_through_kxky} and using unitary properties of $Q$ one can write
\begin{align}
\Bigg(\begin{array}{c}
         k_{u}\\
         k_{v}
    \end{array}\Bigg)^T\cdot
    \Bigg(\begin{array}{c}
        \hat{\mathbf{u}}\\
        \hat{\mathbf{v}}
    \end{array}\Bigg)=
    \Bigg(\begin{array}{c}
         k_{x}\\
         k_{y}
    \end{array}\Bigg)^T\cdot (Q^T)^T\cdot Q^T
    \Bigg(\begin{array}{c}
        \hat{\mathbf{x}}\\
        \hat{\mathbf{y}}
    \end{array}\Bigg)=
    \Bigg(\begin{array}{c}
         k_{x}\\
         k_{y}
    \end{array}\Bigg)^T\cdot Q\cdot Q^T
    \Bigg(\begin{array}{c}
        \hat{\mathbf{x}}\\
        \hat{\mathbf{y}}
    \end{array}\Bigg)=
      \Bigg(\begin{array}{c}
         k_{x}\\
         k_{y}
    \end{array}\Bigg)^T\cdot
    \Bigg(\begin{array}{c}
        \hat{\mathbf{x}}\\
        \hat{\mathbf{y}}
    \end{array}\Bigg),
    \label{eq_property_kukvkxky}
\end{align}
which essentially means that $e^{-i(k_x x+k_y y)}=e^{-i(k_u u+k_v v)}$. Regardless of the choice of basis vector ($\hat{\mathbf{x}}$ and $\hat{\mathbf{y}}$ or $\hat{\mathbf{u}}$ and $\hat{\mathbf{v}}$), to find FW and BW coefficients $A$ and $B$, one can apply temporal boundary conditions that dictate continuity of electric and magnetic flux densities, i.e.  $\_D_{1}\Big|_{t_1-\delta}=\_D_{2}\Big|_{t_1+\delta}$ and $\_B_{1}\Big|_{t_1-\delta}=\_B_{2}\Big|_{t_1+\delta}$. In the rotated frame these equations read $Q^T\cdot\={\epsilon}_1\cdot \_E_1\Big|_{t=t_1-\delta}=\={\epsilon}_2
\_E_2\Big|_{t=t_1+\delta}$ and $\_H_1\Big|_{t=t_1-\delta}=\_H_2\Big|_{t=t_1+\delta}$. Using expressions of the fields in Eqs.~\eqref{eq_fields_1} and~\eqref{eq_fields_2} in addition to property in Eq.~\eqref{eq_property_kukvkxky} one obtains a system of equations with respect to coefficients $A$ and $B$, which can be solved
\begin{align}
    A=\frac{(\omega_1+\omega_2)}{2\omega_1}e^{i\omega_1 t_1},\quad
    B=-\frac{(\omega_1-\omega_2)}{2\omega_1}e^{i\omega_1 t_1}.
    \label{eq_coef_A_B}
\end{align}
\begin{comment}
and expanded as   
\begin{subequations}
\begin{subequations}
\begin{align}
\Bigg(\begin{array}{cc}
        \epsilon_{2xx} & \epsilon_{2xy} \\
         \epsilon_{2yx} & \epsilon_{2yy}
    \end{array}\Bigg)
    \_E_2\Big|_{t=t_1+\delta}&=
    \Bigg(\begin{array}{cc}
        \epsilon_{1xx} & 0 \\
        0 & \epsilon_{1yy}
    \end{array}\Bigg)\_E_1\Big|_{t=t_1-\delta},\\
\_H_2\Big|_{t=t_1+\delta}&=\_H_1\Big|_{t=t_1-\delta},    
\end{align}
\end{subequations}

\begin{align}
\Bigg(\begin{array}{c}
        \epsilon_{2xx}E_{2,x} + \epsilon_{2xy}E_{2,y} \\
         \epsilon_{2yx}E_{2,x} + \epsilon_{2yy}E_{2,y}
    \end{array}\Bigg)
    &=
    \Bigg(\begin{array}{cc}
        \epsilon_{1xx}E_{1,x}\\
        \epsilon_{1yy}E_{1,y}
    \end{array}\Bigg),\label{eq_D_cont}\\
    H_{2,z}&=H_{1,z}.   
\end{align}
\end{subequations}
\end{comment}

\begin{figure}
\centering 
\begin{subfigure}{0.55\textwidth}
\centering
 \includegraphics[width=\textwidth]{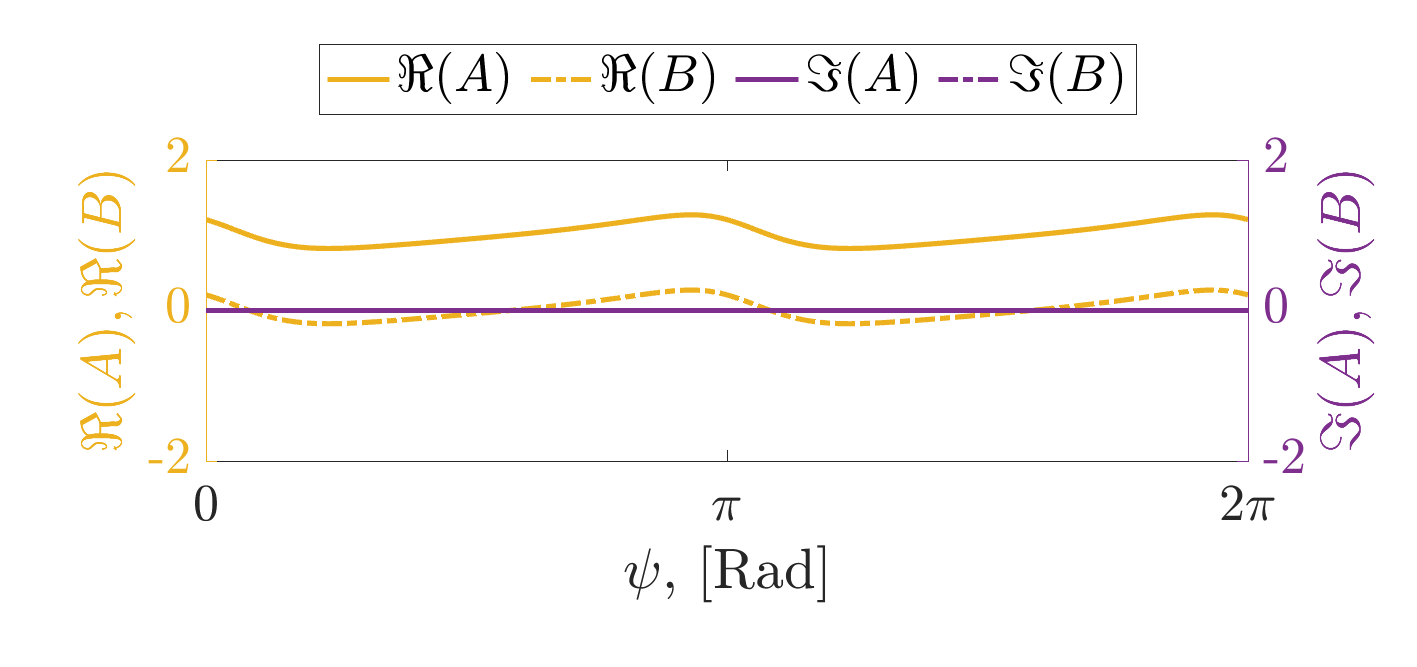}
\caption{}
\end{subfigure}
~
\begin{subfigure}{0.32\textwidth}
\centering
 \includegraphics[width=\textwidth]{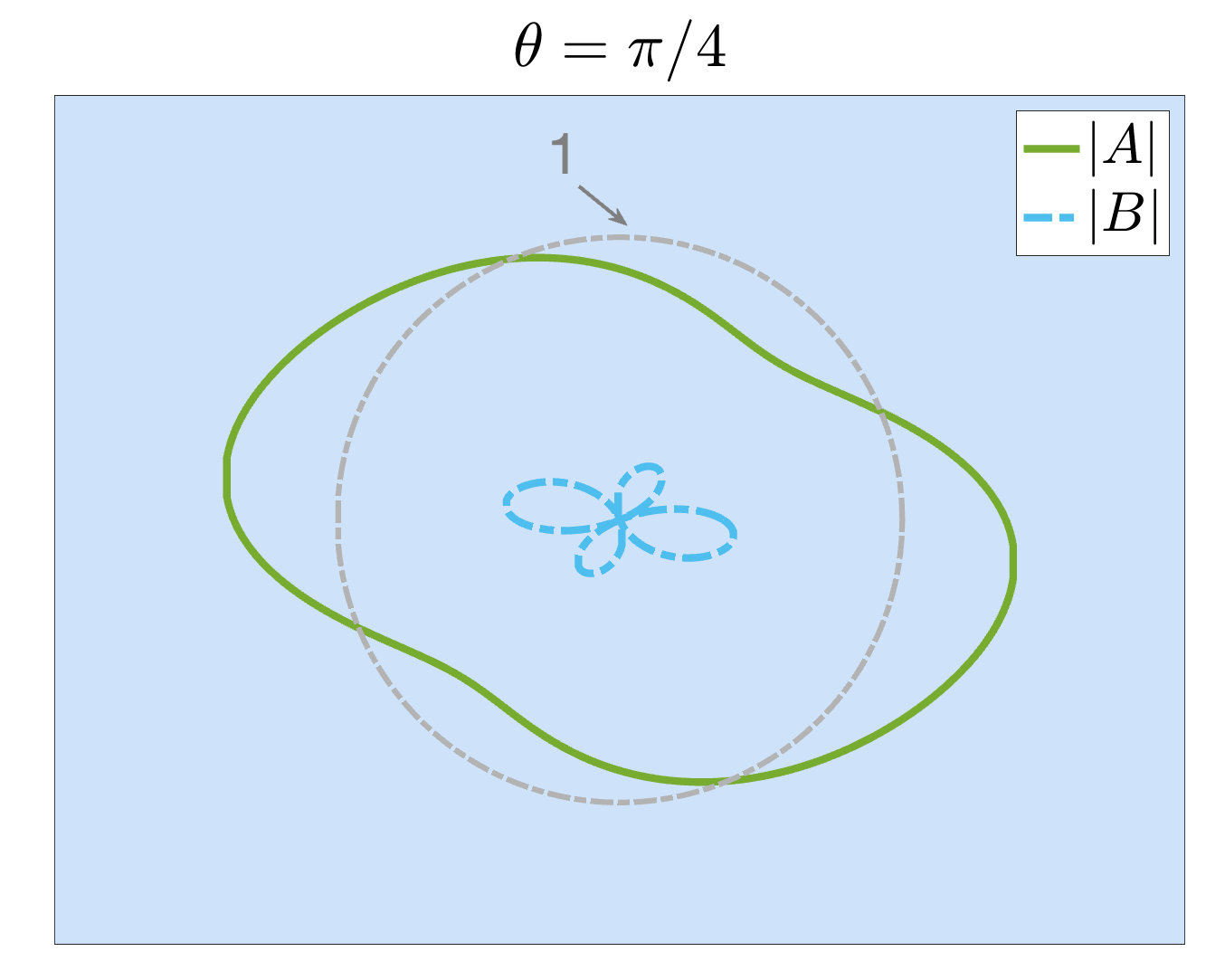}
\caption{}
\end{subfigure}\\
\begin{subfigure}{0.55\textwidth}
\centering
 \includegraphics[width=\textwidth]{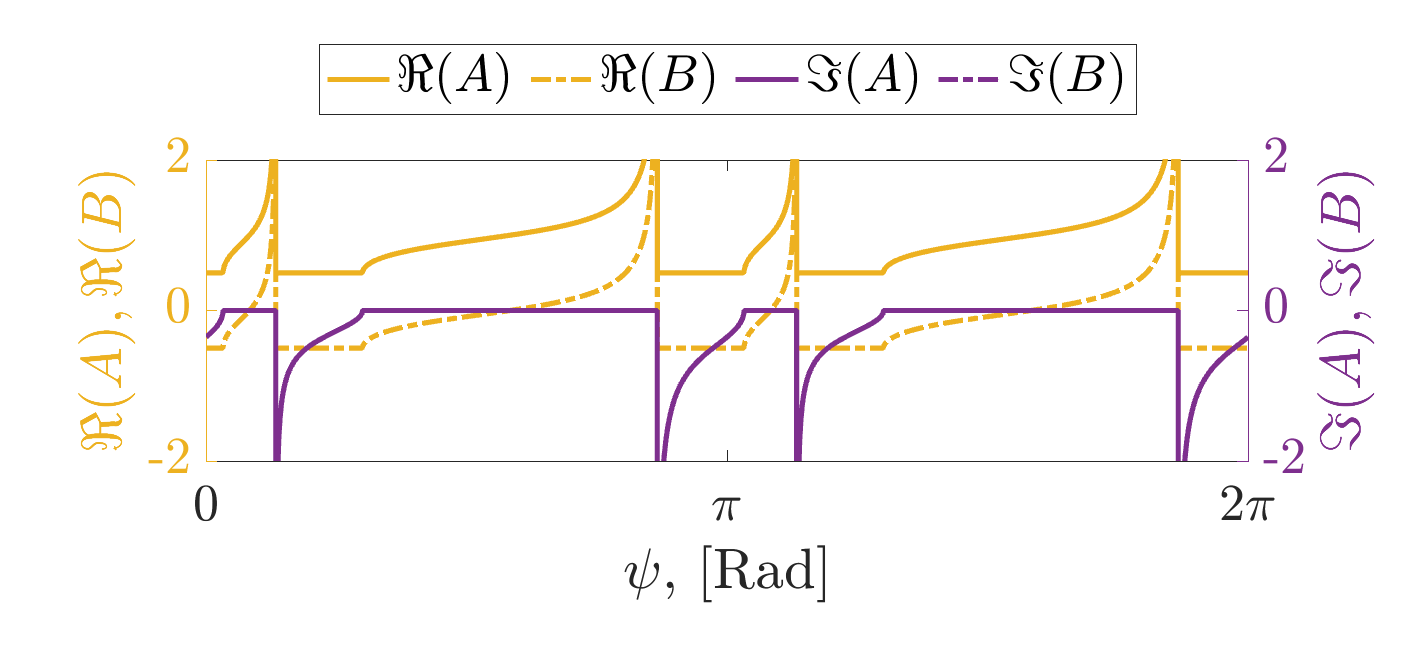}
\caption{}
\end{subfigure}
~
\begin{subfigure}{0.32\textwidth}
\centering
 \includegraphics[width=\textwidth]{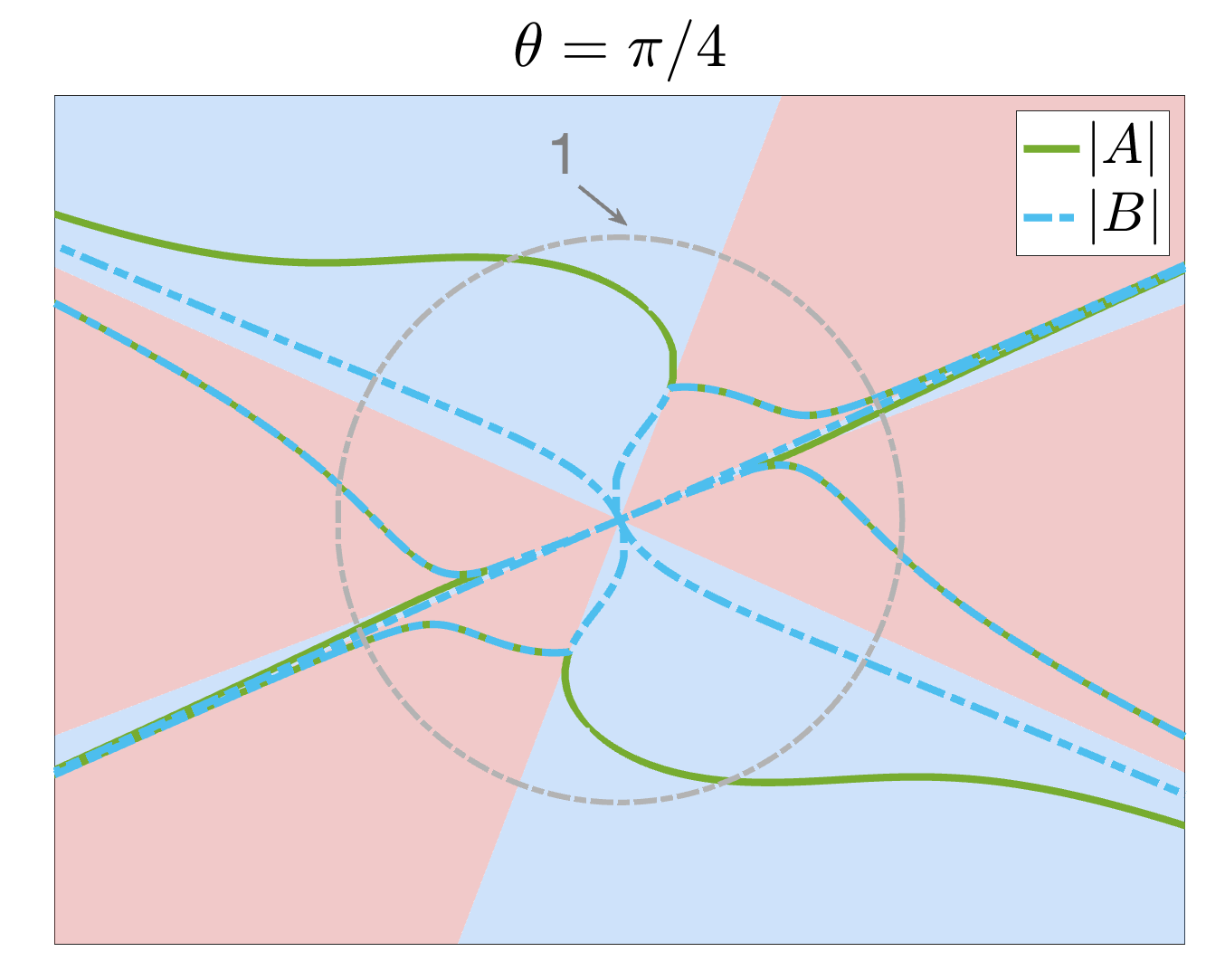}
\caption{}
\end{subfigure}
\caption{(a) and (c) Real and imaginary parts of FW and BW coefficients $A$ and $B$, respectively, after the abrupt rotation-like material change in time at $t_1$=0, with angle $\theta=\pi/4$. (a) Elliptic case with $\epsilon_{1xx}=1$ and $\epsilon_{1yy}=5$. (c) Hyperbolic case with $\epsilon_{1xx}=1$ and $\epsilon_{1yy}=-5$.  We note that the imaginary parts of $A$ and $B$ are the same for all $\psi$'s. (b) and (d) Show absolute values of coefficients $A$ and $B$ in polar coordinates. Blue and red areas, respectively, show regions where $\omega_2$ is real and purely imaginary. The polar angle in these coordinates represent the direction of propagation, $\psi$, of the original plane wave. }
\label{fig_Fields_Amp_phase}
\end{figure}

Figure~\ref{fig_Fields_Amp_phase}(a) and~(c) demonstrates real and imaginary parts of $A$ and $B$ coefficients for both cases under study - elliptic medium ($\epsilon_{1xx}=1$, $\epsilon_{1yy}=5$) and hyperbolic medium ($\epsilon_{1xx}=1$, $\epsilon_{1yy}=-5$) when $t_1$ = 0. For both cases, rotation-like material change is considered with $\theta=\pi/4$. From this figure one can see that, when $t_1$ = 0, imaginary parts of $A$ and $B$ are the same and real parts are different by 1. This can be also seen from  Eq.~\eqref{eq_coef_A_B} by noting that only $\omega_2$ may attain purely imaginary values and $\omega_1$ is set real for all $\psi$. For the elliptic case (i.e., panel (a)), for a certain range of $\psi$, the BW coefficient $B$ may attain positive values, and for some other range of $\psi$, it may be negative.  However, for the hyperbolic case (panel (c)) when $t_1$=0, there are regions of $\psi$ in which the $A$ and $B$  can be complex (and their magnitudes are the same).  These $\psi$'s are those angles of incidence for which the converted frequency $\omega_2$ is purely imaginary.  As mentioned earlier, these are the scenarios in which the original directions
of propagation with frequency $\omega_1$ are along the forbidden (or allowed) directions of hyperbolic medium, and after the temporal change these directions become along the allowed (or forbidden) directions in the rotated hyperbolic material.  It is interesting to note that there are certain $\psi$'s for which the coefficient $B$ is zero, which means that there is no backward wave, $\omega_2=\omega_1$ and $A=1$ for these $\psi$'s (see Fig.~\ref{fig_Fields_Amp_phase}(b) and~(d)).

One can also study the flow of time-average power flux density in this system using Poynting vector $\_S=\frac{1}{2}\Re\big(\_E\times\_H^*$\big). Initially the medium is anisotropic, and therefore $\_S_1\nparallel\textit{\textbf{k}}$.  After the temporal jump the picture becomes even more complex. Using the fields in Eq.~\eqref{eq_fields_2} we get
\begin{subequations}
    \begin{align}    
    S_{2,u}^{+}&=\Re\bigg(\frac{k_u}{\epsilon_{2vv}}\frac{|A|^2}{2\epsilon_0\omega_2}\bigg),\\    
    S_{2,v}^{+}&=\Re\bigg(\frac{k_v}{\epsilon_{2uu}}\frac{|A|^2}{2\epsilon_0\omega_2}\bigg),\\    
    S_{2,u}^{-}&=-\Re\bigg(\frac{k_u}{\epsilon_{2vv}}\frac{|B|^2}{2\epsilon_0\omega_2}\bigg),\\    
    S_{2,v}^{-}&=-\Re\bigg(\frac{k_v}{\epsilon_{2uu}}\frac{|B|^2}{2\epsilon_0\omega_2}\bigg).
    \end{align}
\end{subequations}
These expressions are valid only when $\omega_2$ and $k_x$ and $k_y$ are real quantities.  For those angles of incidence where these quantities are purely imaginary, the FW and BW waves are no longer varying with time sinusoidally, but instead they are growing and decaying with time (under the non-Foster condition due to assumption of no dispersion).  Therefore, the notion of time-average Poynting vector is no longer applicable here.  This is similar to the case of non-Foster temporal change of permittivity from a positive to a negative value studied in Ref.~\cite{pacheco2023holding} in which the notion of space-average (instead of time-average) Poynting vector was considered.

In summary, we have introduced the notion of temporal twistronics or temporal moir\'e in wave propagation in anisotropic media.  Under the dispersionless assumption for the real-valued permittivity tensors, for a monochromatic plane wave propagation in an anisotropic medium we have shown how abrupt temporal rotation-like change in materials parameters can lead to frequency conversion that depends on various factors including the direction of propagation, rotation angle, and initial values of the material parameters.  We have also evaluated the field amplitudes and the time-average Poynting vector after such temporal jump.  This temporal twistronics can offer an interesting path in light-matter interaction in four-dimensional metamaterials.   
%\red{With the known wave vector $\textit{\textbf{k}}$ one obtains the full knowledge of the fields. Figure~\ref{fig_Fields_Amp_phase} demonstrates field amplitudes and phase for $x$ and $y$ components for both cases under study - elliptical medium ($\epsilon_{1xx}=1$, $\epsilon_{1yy}=5$) and hyperbolic medium ($\epsilon_{1xx}=1$, $\epsilon_{1yy}=-5$). For both cases, rotation-like material change is considered with $\theta=\pi/4$. Additionally, one can study the flow of energy in this system using Poynting vector $\_S=\frac{1}{2}\Re\big(\_E\times\_H^*$\big). Initially the medium is anisotropic, therefore $\_S_1\nparallel\textit{\textbf{k}}$, however after the jump the picture becomes even more complex. Figure.~\ref{fig_Poynting_vector} shows Poynting vector before and after an abrupt rotation-like material change with $\theta=\pi/4$ for elliptical and hyperbolic medium. In this figure, wave vector $\textit{\textbf{k}}$ for a given direction $\psi$ is a vector that connects origin with the contour $|\_S|$. Worth noting that for certain $\psi$, the angle between $\textit{\textbf{k}}$ and $\_S_2$ is $\pi/2$ and even more.  }

\begin{acknowledgments}
This work is supported in part by the Ulla Tuominen Foundation to G.P., and in part by the Simons Foundation/Collaboration on Symmetry-Driven Extreme Wave Phenomena (grant 733684) to N.E.  The authors thank Dr. Diego M. Solis of University of Extremadura, Spain, for his valuable comments on the earlier draft of the manuscript.
\end{acknowledgments}

\bibliography{references1}

\begin{thebibliography}{39}%
\makeatletter
\providecommand \@ifxundefined [1]{%
 \@ifx{#1\undefined}
}%
\providecommand \@ifnum [1]{%
 \ifnum #1\expandafter \@firstoftwo
 \else \expandafter \@secondoftwo
 \fi
}%
\providecommand \@ifx [1]{%
 \ifx #1\expandafter \@firstoftwo
 \else \expandafter \@secondoftwo
 \fi
}%
\providecommand \natexlab [1]{#1}%
\providecommand \enquote  [1]{``#1''}%
\providecommand \bibnamefont  [1]{#1}%
\providecommand \bibfnamefont [1]{#1}%
\providecommand \citenamefont [1]{#1}%
\providecommand \href@noop [0]{\@secondoftwo}%
\providecommand \href [0]{\begingroup \@sanitize@url \@href}%
\providecommand \@href[1]{\@@startlink{#1}\@@href}%
\providecommand \@@href[1]{\endgroup#1\@@endlink}%
\providecommand \@sanitize@url [0]{\catcode `\\12\catcode `\$12\catcode
  `\&12\catcode `\#12\catcode `\^12\catcode `\_12\catcode `\%12\relax}%
\providecommand \@@startlink[1]{}%
\providecommand \@@endlink[0]{}%
\providecommand \url  [0]{\begingroup\@sanitize@url \@url }%
\providecommand \@url [1]{\endgroup\@href {#1}{\urlprefix }}%
\providecommand \urlprefix  [0]{URL }%
\providecommand \Eprint [0]{\href }%
\providecommand \doibase [0]{https://doi.org/}%
\providecommand \selectlanguage [0]{\@gobble}%
\providecommand \bibinfo  [0]{\@secondoftwo}%
\providecommand \bibfield  [0]{\@secondoftwo}%
\providecommand \translation [1]{[#1]}%
\providecommand \BibitemOpen [0]{}%
\providecommand \bibitemStop [0]{}%
\providecommand \bibitemNoStop [0]{.\EOS\space}%
\providecommand \EOS [0]{\spacefactor3000\relax}%
\providecommand \BibitemShut  [1]{\csname bibitem#1\endcsname}%
\let\auto@bib@innerbib\@empty
%</preamble>
\bibitem [{\citenamefont {Fresnel}(1821)}]{fresnel1821note}%
  \BibitemOpen
  \bibfield  {author} {\bibinfo {author} {\bibfnamefont {A.}~\bibnamefont
  {Fresnel}},\ }\bibfield  {title} {\bibinfo {title} {Note sur le calcul des
  teintes que la polarisation d{\'e}veloppe dans les lames cristallis{\'e}es},\
  }\href@noop {} {\bibfield  {journal} {\bibinfo  {journal} {Annales de Chimie
  et de Physique}\ }\textbf {\bibinfo {volume} {17}},\ \bibinfo {pages} {101}
  (\bibinfo {year} {1821})}\BibitemShut {NoStop}%
\bibitem [{\citenamefont {Yariv}\ and\ \citenamefont
  {Yeh}(1983)}]{yariv1983optical}%
  \BibitemOpen
  \bibfield  {author} {\bibinfo {author} {\bibfnamefont {A.}~\bibnamefont
  {Yariv}}\ and\ \bibinfo {author} {\bibfnamefont {P.}~\bibnamefont {Yeh}},\
  }\href@noop {} {\emph {\bibinfo {title} {Optical waves in crystal propagation
  and control of laser radiation}}}\ (\bibinfo  {publisher} {John Wiley and
  Sons, Inc., New York, NY},\ \bibinfo {year} {1983})\BibitemShut {NoStop}%
\bibitem [{\citenamefont {Poddubny}\ \emph {et~al.}(2013)\citenamefont
  {Poddubny}, \citenamefont {Iorsh}, \citenamefont {Belov},\ and\ \citenamefont
  {Kivshar}}]{poddubny2013hyperbolic}%
  \BibitemOpen
  \bibfield  {author} {\bibinfo {author} {\bibfnamefont {A.}~\bibnamefont
  {Poddubny}}, \bibinfo {author} {\bibfnamefont {I.}~\bibnamefont {Iorsh}},
  \bibinfo {author} {\bibfnamefont {P.}~\bibnamefont {Belov}},\ and\ \bibinfo
  {author} {\bibfnamefont {Y.}~\bibnamefont {Kivshar}},\ }\bibfield  {title}
  {\bibinfo {title} {Hyperbolic metamaterials},\ }\href@noop {} {\bibfield
  {journal} {\bibinfo  {journal} {Nature photonics}\ }\textbf {\bibinfo
  {volume} {7}},\ \bibinfo {pages} {948} (\bibinfo {year} {2013})}\BibitemShut
  {NoStop}%
\bibitem [{\citenamefont {Ferrari}\ \emph {et~al.}(2015)\citenamefont
  {Ferrari}, \citenamefont {Wu}, \citenamefont {Lepage}, \citenamefont
  {Zhang},\ and\ \citenamefont {Liu}}]{ferrari2015hyperbolic}%
  \BibitemOpen
  \bibfield  {author} {\bibinfo {author} {\bibfnamefont {L.}~\bibnamefont
  {Ferrari}}, \bibinfo {author} {\bibfnamefont {C.}~\bibnamefont {Wu}},
  \bibinfo {author} {\bibfnamefont {D.}~\bibnamefont {Lepage}}, \bibinfo
  {author} {\bibfnamefont {X.}~\bibnamefont {Zhang}},\ and\ \bibinfo {author}
  {\bibfnamefont {Z.}~\bibnamefont {Liu}},\ }\bibfield  {title} {\bibinfo
  {title} {Hyperbolic metamaterials and their applications},\ }\href@noop {}
  {\bibfield  {journal} {\bibinfo  {journal} {Progress in Quantum Electronics}\
  }\textbf {\bibinfo {volume} {40}},\ \bibinfo {pages} {1} (\bibinfo {year}
  {2015})}\BibitemShut {NoStop}%
\bibitem [{\citenamefont {Shekhar}\ \emph {et~al.}(2014)\citenamefont
  {Shekhar}, \citenamefont {Atkinson},\ and\ \citenamefont
  {Jacob}}]{shekhar2014hyperbolic}%
  \BibitemOpen
  \bibfield  {author} {\bibinfo {author} {\bibfnamefont {P.}~\bibnamefont
  {Shekhar}}, \bibinfo {author} {\bibfnamefont {J.}~\bibnamefont {Atkinson}},\
  and\ \bibinfo {author} {\bibfnamefont {Z.}~\bibnamefont {Jacob}},\ }\bibfield
   {title} {\bibinfo {title} {Hyperbolic metamaterials: fundamentals and
  applications},\ }\href@noop {} {\bibfield  {journal} {\bibinfo  {journal}
  {Nano convergence}\ }\textbf {\bibinfo {volume} {1}},\ \bibinfo {pages} {1}
  (\bibinfo {year} {2014})}\BibitemShut {NoStop}%
\bibitem [{\citenamefont {Gomez-Diaz}\ and\ \citenamefont
  {Al\`u}(2016)}]{gomez2016flatland}%
  \BibitemOpen
  \bibfield  {author} {\bibinfo {author} {\bibfnamefont {J.}~\bibnamefont
  {Gomez-Diaz}}\ and\ \bibinfo {author} {\bibfnamefont {A.}~\bibnamefont
  {Al\`u}},\ }\bibfield  {title} {\bibinfo {title} {Flatland optics with
  hyperbolic metasurfaces},\ }\href@noop {} {\bibfield  {journal} {\bibinfo
  {journal} {ACS Photonics}\ }\textbf {\bibinfo {volume} {3}},\ \bibinfo
  {pages} {2211} (\bibinfo {year} {2016})}\BibitemShut {NoStop}%
\bibitem [{\citenamefont {High}\ \emph {et~al.}(2015)\citenamefont {High},
  \citenamefont {Devlin}, \citenamefont {Dibos}, \citenamefont {Polking},
  \citenamefont {Wild}, \citenamefont {Perczel}, \citenamefont {De~Leon},
  \citenamefont {Lukin},\ and\ \citenamefont {Park}}]{high2015visible}%
  \BibitemOpen
  \bibfield  {author} {\bibinfo {author} {\bibfnamefont {A.~A.}\ \bibnamefont
  {High}}, \bibinfo {author} {\bibfnamefont {R.~C.}\ \bibnamefont {Devlin}},
  \bibinfo {author} {\bibfnamefont {A.}~\bibnamefont {Dibos}}, \bibinfo
  {author} {\bibfnamefont {M.}~\bibnamefont {Polking}}, \bibinfo {author}
  {\bibfnamefont {D.~S.}\ \bibnamefont {Wild}}, \bibinfo {author}
  {\bibfnamefont {J.}~\bibnamefont {Perczel}}, \bibinfo {author} {\bibfnamefont
  {N.~P.}\ \bibnamefont {De~Leon}}, \bibinfo {author} {\bibfnamefont {M.~D.}\
  \bibnamefont {Lukin}},\ and\ \bibinfo {author} {\bibfnamefont
  {H.}~\bibnamefont {Park}},\ }\bibfield  {title} {\bibinfo {title}
  {Visible-frequency hyperbolic metasurface},\ }\href@noop {} {\bibfield
  {journal} {\bibinfo  {journal} {Nature}\ }\textbf {\bibinfo {volume} {522}},\
  \bibinfo {pages} {192} (\bibinfo {year} {2015})}\BibitemShut {NoStop}%
\bibitem [{\citenamefont {Huo}\ \emph {et~al.}(2019)\citenamefont {Huo},
  \citenamefont {Zhang}, \citenamefont {Liang}, \citenamefont {Lu},\ and\
  \citenamefont {Xu}}]{huo2019hyperbolic}%
  \BibitemOpen
  \bibfield  {author} {\bibinfo {author} {\bibfnamefont {P.}~\bibnamefont
  {Huo}}, \bibinfo {author} {\bibfnamefont {S.}~\bibnamefont {Zhang}}, \bibinfo
  {author} {\bibfnamefont {Y.}~\bibnamefont {Liang}}, \bibinfo {author}
  {\bibfnamefont {Y.}~\bibnamefont {Lu}},\ and\ \bibinfo {author}
  {\bibfnamefont {T.}~\bibnamefont {Xu}},\ }\bibfield  {title} {\bibinfo
  {title} {Hyperbolic metamaterials and metasurfaces: fundamentals and
  applications},\ }\href@noop {} {\bibfield  {journal} {\bibinfo  {journal}
  {Advanced Optical Materials}\ }\textbf {\bibinfo {volume} {7}},\ \bibinfo
  {pages} {1801616} (\bibinfo {year} {2019})}\BibitemShut {NoStop}%
\bibitem [{\citenamefont {Guo}\ \emph {et~al.}(2020)\citenamefont {Guo},
  \citenamefont {Jiang},\ and\ \citenamefont {Chen}}]{guo2020hyperbolic}%
  \BibitemOpen
  \bibfield  {author} {\bibinfo {author} {\bibfnamefont {Z.}~\bibnamefont
  {Guo}}, \bibinfo {author} {\bibfnamefont {H.}~\bibnamefont {Jiang}},\ and\
  \bibinfo {author} {\bibfnamefont {H.}~\bibnamefont {Chen}},\ }\bibfield
  {title} {\bibinfo {title} {Hyperbolic metamaterials: From dispersion
  manipulation to applications},\ }\href@noop {} {\bibfield  {journal}
  {\bibinfo  {journal} {Journal of Applied Physics}\ }\textbf {\bibinfo
  {volume} {127}} (\bibinfo {year} {2020})}\BibitemShut {NoStop}%
\bibitem [{\citenamefont {Carr}\ \emph {et~al.}(2017)\citenamefont {Carr},
  \citenamefont {Massatt}, \citenamefont {Fang}, \citenamefont {Cazeaux},
  \citenamefont {Luskin},\ and\ \citenamefont {Kaxiras}}]{carr2017twistronics}%
  \BibitemOpen
  \bibfield  {author} {\bibinfo {author} {\bibfnamefont {S.}~\bibnamefont
  {Carr}}, \bibinfo {author} {\bibfnamefont {D.}~\bibnamefont {Massatt}},
  \bibinfo {author} {\bibfnamefont {S.}~\bibnamefont {Fang}}, \bibinfo {author}
  {\bibfnamefont {P.}~\bibnamefont {Cazeaux}}, \bibinfo {author} {\bibfnamefont
  {M.}~\bibnamefont {Luskin}},\ and\ \bibinfo {author} {\bibfnamefont
  {E.}~\bibnamefont {Kaxiras}},\ }\bibfield  {title} {\bibinfo {title}
  {Twistronics: Manipulating the electronic properties of two-dimensional
  layered structures through their twist angle},\ }\href@noop {} {\bibfield
  {journal} {\bibinfo  {journal} {Physical Review B}\ }\textbf {\bibinfo
  {volume} {95}},\ \bibinfo {pages} {075420} (\bibinfo {year}
  {2017})}\BibitemShut {NoStop}%
\bibitem [{\citenamefont {Ren}\ \emph {et~al.}(2020)\citenamefont {Ren},
  \citenamefont {Zhang}, \citenamefont {Liu},\ and\ \citenamefont
  {He}}]{ren2020twistronics}%
  \BibitemOpen
  \bibfield  {author} {\bibinfo {author} {\bibfnamefont {Y.-N.}\ \bibnamefont
  {Ren}}, \bibinfo {author} {\bibfnamefont {Y.}~\bibnamefont {Zhang}}, \bibinfo
  {author} {\bibfnamefont {Y.-W.}\ \bibnamefont {Liu}},\ and\ \bibinfo {author}
  {\bibfnamefont {L.}~\bibnamefont {He}},\ }\bibfield  {title} {\bibinfo
  {title} {Twistronics in graphene-based van der waals structures},\
  }\href@noop {} {\bibfield  {journal} {\bibinfo  {journal} {Chinese Physics
  B}\ }\textbf {\bibinfo {volume} {29}},\ \bibinfo {pages} {117303} (\bibinfo
  {year} {2020})}\BibitemShut {NoStop}%
\bibitem [{\citenamefont {Hu}\ \emph {et~al.}(2021)\citenamefont {Hu},
  \citenamefont {Qiu},\ and\ \citenamefont {Al{\`u}}}]{hu2021twistronics}%
  \BibitemOpen
  \bibfield  {author} {\bibinfo {author} {\bibfnamefont {G.}~\bibnamefont
  {Hu}}, \bibinfo {author} {\bibfnamefont {C.-W.}\ \bibnamefont {Qiu}},\ and\
  \bibinfo {author} {\bibfnamefont {A.}~\bibnamefont {Al{\`u}}},\ }\bibfield
  {title} {\bibinfo {title} {Twistronics for photons: opinion},\ }\href@noop {}
  {\bibfield  {journal} {\bibinfo  {journal} {Optical Materials Express}\
  }\textbf {\bibinfo {volume} {11}},\ \bibinfo {pages} {1377} (\bibinfo {year}
  {2021})}\BibitemShut {NoStop}%
\bibitem [{\citenamefont {Hu}\ \emph {et~al.}(2020)\citenamefont {Hu},
  \citenamefont {Ou}, \citenamefont {Si}, \citenamefont {Wu}, \citenamefont
  {Wu}, \citenamefont {Dai}, \citenamefont {Krasnok}, \citenamefont {Mazor},
  \citenamefont {Zhang}, \citenamefont {Bao} \emph
  {et~al.}}]{hu2020topological}%
  \BibitemOpen
  \bibfield  {author} {\bibinfo {author} {\bibfnamefont {G.}~\bibnamefont
  {Hu}}, \bibinfo {author} {\bibfnamefont {Q.}~\bibnamefont {Ou}}, \bibinfo
  {author} {\bibfnamefont {G.}~\bibnamefont {Si}}, \bibinfo {author}
  {\bibfnamefont {Y.}~\bibnamefont {Wu}}, \bibinfo {author} {\bibfnamefont
  {J.}~\bibnamefont {Wu}}, \bibinfo {author} {\bibfnamefont {Z.}~\bibnamefont
  {Dai}}, \bibinfo {author} {\bibfnamefont {A.}~\bibnamefont {Krasnok}},
  \bibinfo {author} {\bibfnamefont {Y.}~\bibnamefont {Mazor}}, \bibinfo
  {author} {\bibfnamefont {Q.}~\bibnamefont {Zhang}}, \bibinfo {author}
  {\bibfnamefont {Q.}~\bibnamefont {Bao}}, \emph {et~al.},\ }\bibfield  {title}
  {\bibinfo {title} {Topological polaritons and photonic magic angles in
  twisted $\alpha$-{M}o{O}$_3$ bilayers},\ }\href@noop {} {\bibfield  {journal}
  {\bibinfo  {journal} {Nature}\ }\textbf {\bibinfo {volume} {582}},\ \bibinfo
  {pages} {209} (\bibinfo {year} {2020})}\BibitemShut {NoStop}%
\bibitem [{\citenamefont {Tang}\ \emph {et~al.}(2023)\citenamefont {Tang},
  \citenamefont {Lou}, \citenamefont {Du}, \citenamefont {Zhang}, \citenamefont
  {Ni}, \citenamefont {Xu}, \citenamefont {Jin}, \citenamefont {Fan},\ and\
  \citenamefont {Mazur}}]{tang2023experimental}%
  \BibitemOpen
  \bibfield  {author} {\bibinfo {author} {\bibfnamefont {H.}~\bibnamefont
  {Tang}}, \bibinfo {author} {\bibfnamefont {B.}~\bibnamefont {Lou}}, \bibinfo
  {author} {\bibfnamefont {F.}~\bibnamefont {Du}}, \bibinfo {author}
  {\bibfnamefont {M.}~\bibnamefont {Zhang}}, \bibinfo {author} {\bibfnamefont
  {X.}~\bibnamefont {Ni}}, \bibinfo {author} {\bibfnamefont {W.}~\bibnamefont
  {Xu}}, \bibinfo {author} {\bibfnamefont {R.}~\bibnamefont {Jin}}, \bibinfo
  {author} {\bibfnamefont {S.}~\bibnamefont {Fan}},\ and\ \bibinfo {author}
  {\bibfnamefont {E.}~\bibnamefont {Mazur}},\ }\bibfield  {title} {\bibinfo
  {title} {Experimental probe of twist angle--dependent band structure of
  on-chip optical bilayer photonic crystal},\ }\href@noop {} {\bibfield
  {journal} {\bibinfo  {journal} {Science Advances}\ }\textbf {\bibinfo
  {volume} {9}},\ \bibinfo {pages} {eadh8498} (\bibinfo {year}
  {2023})}\BibitemShut {NoStop}%
\bibitem [{\citenamefont {Duan}\ \emph {et~al.}(2020)\citenamefont {Duan},
  \citenamefont {Capote-Robayna}, \citenamefont {Taboada-Guti{\'e}rrez},
  \citenamefont {{\'A}lvarez-P{\'e}rez}, \citenamefont {Prieto}, \citenamefont
  {Mart{\'\i}n-S{\'a}nchez}, \citenamefont {Nikitin},\ and\ \citenamefont
  {Alonso-Gonz{\'a}lez}}]{duan2020twisted}%
  \BibitemOpen
  \bibfield  {author} {\bibinfo {author} {\bibfnamefont {J.}~\bibnamefont
  {Duan}}, \bibinfo {author} {\bibfnamefont {N.}~\bibnamefont
  {Capote-Robayna}}, \bibinfo {author} {\bibfnamefont {J.}~\bibnamefont
  {Taboada-Guti{\'e}rrez}}, \bibinfo {author} {\bibfnamefont {G.}~\bibnamefont
  {{\'A}lvarez-P{\'e}rez}}, \bibinfo {author} {\bibfnamefont {I.}~\bibnamefont
  {Prieto}}, \bibinfo {author} {\bibfnamefont {J.}~\bibnamefont
  {Mart{\'\i}n-S{\'a}nchez}}, \bibinfo {author} {\bibfnamefont {A.~Y.}\
  \bibnamefont {Nikitin}},\ and\ \bibinfo {author} {\bibfnamefont
  {P.}~\bibnamefont {Alonso-Gonz{\'a}lez}},\ }\bibfield  {title} {\bibinfo
  {title} {Twisted nano-optics: manipulating light at the nanoscale with
  twisted phonon polaritonic slabs},\ }\href@noop {} {\bibfield  {journal}
  {\bibinfo  {journal} {Nano Letters}\ }\textbf {\bibinfo {volume} {20}},\
  \bibinfo {pages} {5323} (\bibinfo {year} {2020})}\BibitemShut {NoStop}%
\bibitem [{\citenamefont {Ciarrocchi}\ \emph {et~al.}(2022)\citenamefont
  {Ciarrocchi}, \citenamefont {Tagarelli}, \citenamefont {Avsar},\ and\
  \citenamefont {Kis}}]{ciarrocchi2022excitonic}%
  \BibitemOpen
  \bibfield  {author} {\bibinfo {author} {\bibfnamefont {A.}~\bibnamefont
  {Ciarrocchi}}, \bibinfo {author} {\bibfnamefont {F.}~\bibnamefont
  {Tagarelli}}, \bibinfo {author} {\bibfnamefont {A.}~\bibnamefont {Avsar}},\
  and\ \bibinfo {author} {\bibfnamefont {A.}~\bibnamefont {Kis}},\ }\bibfield
  {title} {\bibinfo {title} {Excitonic devices with van der waals
  heterostructures: valleytronics meets twistronics},\ }\href@noop {}
  {\bibfield  {journal} {\bibinfo  {journal} {Nature Reviews Materials}\
  }\textbf {\bibinfo {volume} {7}},\ \bibinfo {pages} {449} (\bibinfo {year}
  {2022})}\BibitemShut {NoStop}%
\bibitem [{\citenamefont {Zadeh}(1950{\natexlab{a}})}]{zadeh1950frequency}%
  \BibitemOpen
  \bibfield  {author} {\bibinfo {author} {\bibfnamefont {L.~A.}\ \bibnamefont
  {Zadeh}},\ }\bibfield  {title} {\bibinfo {title} {Frequency analysis of
  variable networks},\ }\href@noop {} {\bibfield  {journal} {\bibinfo
  {journal} {Proceedings of the IRE}\ }\textbf {\bibinfo {volume} {38}},\
  \bibinfo {pages} {291} (\bibinfo {year} {1950}{\natexlab{a}})}\BibitemShut
  {NoStop}%
\bibitem [{\citenamefont {Zadeh}(1950{\natexlab{b}})}]{zadeh1950determination}%
  \BibitemOpen
  \bibfield  {author} {\bibinfo {author} {\bibfnamefont {L.~A.}\ \bibnamefont
  {Zadeh}},\ }\bibfield  {title} {\bibinfo {title} {The determination of the
  impulsive response of variable networks},\ }\href@noop {} {\bibfield
  {journal} {\bibinfo  {journal} {Journal of Applied Physics}\ }\textbf
  {\bibinfo {volume} {21}},\ \bibinfo {pages} {642} (\bibinfo {year}
  {1950}{\natexlab{b}})}\BibitemShut {NoStop}%
\bibitem [{\citenamefont {Morgenthaler}(1958)}]{morgenthaler1958velocity}%
  \BibitemOpen
  \bibfield  {author} {\bibinfo {author} {\bibfnamefont {F.~R.}\ \bibnamefont
  {Morgenthaler}},\ }\bibfield  {title} {\bibinfo {title} {Velocity modulation
  of electromagnetic waves},\ }\href@noop {} {\bibfield  {journal} {\bibinfo
  {journal} {IRE Transactions on Microwave Theory and Techniques}\ }\textbf
  {\bibinfo {volume} {6}},\ \bibinfo {pages} {167} (\bibinfo {year}
  {1958})}\BibitemShut {NoStop}%
\bibitem [{\citenamefont {Cullen}(1958)}]{cullen1958travelling}%
  \BibitemOpen
  \bibfield  {author} {\bibinfo {author} {\bibfnamefont {A.}~\bibnamefont
  {Cullen}},\ }\bibfield  {title} {\bibinfo {title} {A travelling-wave
  parametric amplifier},\ }\href@noop {} {\bibfield  {journal} {\bibinfo
  {journal} {Nature}\ }\textbf {\bibinfo {volume} {181}},\ \bibinfo {pages}
  {332} (\bibinfo {year} {1958})}\BibitemShut {NoStop}%
\bibitem [{\citenamefont {Tien}(1958)}]{tien1958parametric}%
  \BibitemOpen
  \bibfield  {author} {\bibinfo {author} {\bibfnamefont {P.}~\bibnamefont
  {Tien}},\ }\bibfield  {title} {\bibinfo {title} {Parametric amplification and
  frequency mixing in propagating circuits},\ }\href@noop {} {\bibfield
  {journal} {\bibinfo  {journal} {Journal of Applied Physics}\ }\textbf
  {\bibinfo {volume} {29}},\ \bibinfo {pages} {1347} (\bibinfo {year}
  {1958})}\BibitemShut {NoStop}%
\bibitem [{\citenamefont {Engheta}(2020)}]{engheta2020metamaterials}%
  \BibitemOpen
  \bibfield  {author} {\bibinfo {author} {\bibfnamefont {N.}~\bibnamefont
  {Engheta}},\ }\bibfield  {title} {\bibinfo {title} {Metamaterials with high
  degrees of freedom: space, time, and more},\ }\href@noop {} {\bibfield
  {journal} {\bibinfo  {journal} {Nanophotonics}\ }\textbf {\bibinfo {volume}
  {10}},\ \bibinfo {pages} {639} (\bibinfo {year} {2020})}\BibitemShut
  {NoStop}%
\bibitem [{\citenamefont {Engheta}(2023)}]{engheta2023four}%
  \BibitemOpen
  \bibfield  {author} {\bibinfo {author} {\bibfnamefont {N.}~\bibnamefont
  {Engheta}},\ }\bibfield  {title} {\bibinfo {title} {Four-dimensional optics
  using time-varying metamaterials},\ }\href@noop {} {\bibfield  {journal}
  {\bibinfo  {journal} {Science}\ }\textbf {\bibinfo {volume} {379}},\ \bibinfo
  {pages} {1190} (\bibinfo {year} {2023})}\BibitemShut {NoStop}%
\bibitem [{\citenamefont {Tirole}\ \emph {et~al.}(2023)\citenamefont {Tirole},
  \citenamefont {Vezzoli}, \citenamefont {Galiffi}, \citenamefont {Robertson},
  \citenamefont {Maurice}, \citenamefont {Tilmann}, \citenamefont {Maier},
  \citenamefont {Pendry},\ and\ \citenamefont {Sapienza}}]{tirole2023double}%
  \BibitemOpen
  \bibfield  {author} {\bibinfo {author} {\bibfnamefont {R.}~\bibnamefont
  {Tirole}}, \bibinfo {author} {\bibfnamefont {S.}~\bibnamefont {Vezzoli}},
  \bibinfo {author} {\bibfnamefont {E.}~\bibnamefont {Galiffi}}, \bibinfo
  {author} {\bibfnamefont {I.}~\bibnamefont {Robertson}}, \bibinfo {author}
  {\bibfnamefont {D.}~\bibnamefont {Maurice}}, \bibinfo {author} {\bibfnamefont
  {B.}~\bibnamefont {Tilmann}}, \bibinfo {author} {\bibfnamefont {S.~A.}\
  \bibnamefont {Maier}}, \bibinfo {author} {\bibfnamefont {J.~B.}\ \bibnamefont
  {Pendry}},\ and\ \bibinfo {author} {\bibfnamefont {R.}~\bibnamefont
  {Sapienza}},\ }\bibfield  {title} {\bibinfo {title} {Double-slit time
  diffraction at optical frequencies},\ }\href@noop {} {\bibfield  {journal}
  {\bibinfo  {journal} {Nature Physics}\ ,\ \bibinfo {pages} {1}} (\bibinfo
  {year} {2023})}\BibitemShut {NoStop}%
\bibitem [{\citenamefont {Galiffi}\ \emph {et~al.}(2022)\citenamefont
  {Galiffi}, \citenamefont {Tirole}, \citenamefont {Yin}, \citenamefont {Li},
  \citenamefont {Vezzoli}, \citenamefont {Huidobro}, \citenamefont
  {Silveirinha}, \citenamefont {Sapienza}, \citenamefont {Al{\`u}},\ and\
  \citenamefont {Pendry}}]{galiffi_photonics_2022}%
  \BibitemOpen
  \bibfield  {author} {\bibinfo {author} {\bibfnamefont {E.}~\bibnamefont
  {Galiffi}}, \bibinfo {author} {\bibfnamefont {R.}~\bibnamefont {Tirole}},
  \bibinfo {author} {\bibfnamefont {S.}~\bibnamefont {Yin}}, \bibinfo {author}
  {\bibfnamefont {H.}~\bibnamefont {Li}}, \bibinfo {author} {\bibfnamefont
  {S.}~\bibnamefont {Vezzoli}}, \bibinfo {author} {\bibfnamefont {P.~A.}\
  \bibnamefont {Huidobro}}, \bibinfo {author} {\bibfnamefont {M.~G.}\
  \bibnamefont {Silveirinha}}, \bibinfo {author} {\bibfnamefont
  {R.}~\bibnamefont {Sapienza}}, \bibinfo {author} {\bibfnamefont
  {A.}~\bibnamefont {Al{\`u}}},\ and\ \bibinfo {author} {\bibfnamefont {J.~B.}\
  \bibnamefont {Pendry}},\ }\bibfield  {title} {\bibinfo {title} {Photonics of
  time-varying media},\ }\href {https://doi.org/10.1117/1.AP.4.1.014002}
  {\bibfield  {journal} {\bibinfo  {journal} {Advanced Photonics}\ }\textbf
  {\bibinfo {volume} {4}},\ \bibinfo {pages} {014002} (\bibinfo {year}
  {2022})}\BibitemShut {NoStop}%
\bibitem [{\citenamefont {Zhou}\ \emph {et~al.}(2020)\citenamefont {Zhou},
  \citenamefont {Alam}, \citenamefont {Karimi}, \citenamefont {Upham},
  \citenamefont {Reshef}, \citenamefont {Liu}, \citenamefont {Willner},\ and\
  \citenamefont {Boyd}}]{zhou2020broadband}%
  \BibitemOpen
  \bibfield  {author} {\bibinfo {author} {\bibfnamefont {Y.}~\bibnamefont
  {Zhou}}, \bibinfo {author} {\bibfnamefont {M.~Z.}\ \bibnamefont {Alam}},
  \bibinfo {author} {\bibfnamefont {M.}~\bibnamefont {Karimi}}, \bibinfo
  {author} {\bibfnamefont {J.}~\bibnamefont {Upham}}, \bibinfo {author}
  {\bibfnamefont {O.}~\bibnamefont {Reshef}}, \bibinfo {author} {\bibfnamefont
  {C.}~\bibnamefont {Liu}}, \bibinfo {author} {\bibfnamefont {A.~E.}\
  \bibnamefont {Willner}},\ and\ \bibinfo {author} {\bibfnamefont {R.~W.}\
  \bibnamefont {Boyd}},\ }\bibfield  {title} {\bibinfo {title} {Broadband
  frequency translation through time refraction in an epsilon-near-zero
  material},\ }\href@noop {} {\bibfield  {journal} {\bibinfo  {journal} {Nature
  Communications}\ }\textbf {\bibinfo {volume} {11}},\ \bibinfo {pages} {2180}
  (\bibinfo {year} {2020})}\BibitemShut {NoStop}%
\bibitem [{\citenamefont {Pacheco-Pe{\~n}a}\ and\ \citenamefont
  {Engheta}(2020)}]{pacheco2020temporal}%
  \BibitemOpen
  \bibfield  {author} {\bibinfo {author} {\bibfnamefont {V.}~\bibnamefont
  {Pacheco-Pe{\~n}a}}\ and\ \bibinfo {author} {\bibfnamefont {N.}~\bibnamefont
  {Engheta}},\ }\bibfield  {title} {\bibinfo {title} {Temporal aiming},\
  }\href@noop {} {\bibfield  {journal} {\bibinfo  {journal} {Light: Science \&
  Applications}\ }\textbf {\bibinfo {volume} {9}},\ \bibinfo {pages} {129}
  (\bibinfo {year} {2020})}\BibitemShut {NoStop}%
\bibitem [{\citenamefont {Qui{\~n}ones}\ \emph {et~al.}(2021)\citenamefont
  {Qui{\~n}ones}, \citenamefont {Underwood},\ and\ \citenamefont
  {Cappelli}}]{quinones_tunable_2021}%
  \BibitemOpen
  \bibfield  {author} {\bibinfo {author} {\bibfnamefont {R.~A.~C.}\
  \bibnamefont {Qui{\~n}ones}}, \bibinfo {author} {\bibfnamefont {T.~C.}\
  \bibnamefont {Underwood}},\ and\ \bibinfo {author} {\bibfnamefont {M.~A.}\
  \bibnamefont {Cappelli}},\ }\bibfield  {title} {\bibinfo {title} {Tunable
  surface plasmon resonance in laser-induced plasma spheroids},\ }\href
  {https://doi.org/10.1088/1361-6595/abc5a2} {\bibfield  {journal} {\bibinfo
  {journal} {Plasma Sources Science and Technology}\ }\textbf {\bibinfo
  {volume} {30}},\ \bibinfo {pages} {045010} (\bibinfo {year}
  {2021})}\BibitemShut {NoStop}%
\bibitem [{\citenamefont {Yin}\ and\ \citenamefont
  {Al{\`u}}(2022)}]{yin2022efficient}%
  \BibitemOpen
  \bibfield  {author} {\bibinfo {author} {\bibfnamefont {S.}~\bibnamefont
  {Yin}}\ and\ \bibinfo {author} {\bibfnamefont {A.}~\bibnamefont {Al{\`u}}},\
  }\bibfield  {title} {\bibinfo {title} {Efficient phase conjugation in a
  space-time leaky waveguide},\ }\href@noop {} {\bibfield  {journal} {\bibinfo
  {journal} {ACS Photonics}\ }\textbf {\bibinfo {volume} {9}},\ \bibinfo
  {pages} {979} (\bibinfo {year} {2022})}\BibitemShut {NoStop}%
\bibitem [{\citenamefont {Biancalana}\ \emph {et~al.}(2007)\citenamefont
  {Biancalana}, \citenamefont {Amann}, \citenamefont {Uskov},\ and\
  \citenamefont {O'Reilly}}]{biancalana_dynamics_2007}%
  \BibitemOpen
  \bibfield  {author} {\bibinfo {author} {\bibfnamefont {F.}~\bibnamefont
  {Biancalana}}, \bibinfo {author} {\bibfnamefont {A.}~\bibnamefont {Amann}},
  \bibinfo {author} {\bibfnamefont {A.~V.}\ \bibnamefont {Uskov}},\ and\
  \bibinfo {author} {\bibfnamefont {E.~P.}\ \bibnamefont {O'Reilly}},\
  }\bibfield  {title} {\bibinfo {title} {Dynamics of light propagation in
  spatiotemporal dielectric structures},\ }\href
  {https://doi.org/10.1103/PhysRevE.75.046607} {\bibfield  {journal} {\bibinfo
  {journal} {Physical Review E}\ }\textbf {\bibinfo {volume} {75}},\ \bibinfo
  {pages} {046607} (\bibinfo {year} {2007})}\BibitemShut {NoStop}%
\bibitem [{\citenamefont {Zurita-S\'{a}nchez}\ \emph
  {et~al.}(2009)\citenamefont {Zurita-S\'{a}nchez}, \citenamefont {Halevi},\
  and\ \citenamefont {Cervantes-Gonzalez}}]{zurita-sanchez_reflection_2009}%
  \BibitemOpen
  \bibfield  {author} {\bibinfo {author} {\bibfnamefont {J.~R.}\ \bibnamefont
  {Zurita-S\'{a}nchez}}, \bibinfo {author} {\bibfnamefont {P.}~\bibnamefont
  {Halevi}},\ and\ \bibinfo {author} {\bibfnamefont {J.~C.}\ \bibnamefont
  {Cervantes-Gonzalez}},\ }\bibfield  {title} {\bibinfo {title} {Reflection and
  transmission of a wave incident on a slab with a time-periodic dielectric
  function $\varepsilon(t)$},\ }\href
  {https://doi.org/10.1103/PhysRevA.79.053821} {\bibfield  {journal} {\bibinfo
  {journal} {Physical Review A}\ }\textbf {\bibinfo {volume} {79}},\ \bibinfo
  {pages} {053821} (\bibinfo {year} {2009})}\BibitemShut {NoStop}%
\bibitem [{\citenamefont {Reyes-Ayona}\ and\ \citenamefont
  {Halevi}(2015)}]{reyes-ayona_observation_2015}%
  \BibitemOpen
  \bibfield  {author} {\bibinfo {author} {\bibfnamefont {J.~R.}\ \bibnamefont
  {Reyes-Ayona}}\ and\ \bibinfo {author} {\bibfnamefont {P.}~\bibnamefont
  {Halevi}},\ }\bibfield  {title} {\bibinfo {title} {Observation of genuine
  wave vector (k or $\beta$) gap in a dynamic transmission line and temporal
  photonic crystals},\ }\href {https://doi.org/10.1063/1.4928659} {\bibfield
  {journal} {\bibinfo  {journal} {Applied Physics Letters}\ }\textbf {\bibinfo
  {volume} {107}},\ \bibinfo {pages} {074101} (\bibinfo {year}
  {2015})}\BibitemShut {NoStop}%
\bibitem [{\citenamefont {Lustig}\ \emph {et~al.}(2018)\citenamefont {Lustig},
  \citenamefont {Sharabi},\ and\ \citenamefont
  {Segev}}]{lustig_topological_2018}%
  \BibitemOpen
  \bibfield  {author} {\bibinfo {author} {\bibfnamefont {E.}~\bibnamefont
  {Lustig}}, \bibinfo {author} {\bibfnamefont {Y.}~\bibnamefont {Sharabi}},\
  and\ \bibinfo {author} {\bibfnamefont {M.}~\bibnamefont {Segev}},\ }\bibfield
   {title} {\bibinfo {title} {Topological aspects of photonic time crystals},\
  }\href
  {https://www.osapublishing.org/optica/abstract.cfm?uri=optica-5-11-1390}
  {\bibfield  {journal} {\bibinfo  {journal} {Optica}\ }\textbf {\bibinfo
  {volume} {5}},\ \bibinfo {pages} {1390} (\bibinfo {year} {2018})}\BibitemShut
  {NoStop}%
\bibitem [{\citenamefont {Park}\ and\ \citenamefont
  {Min}(2021)}]{park_spatiotemporal_2021}%
  \BibitemOpen
  \bibfield  {author} {\bibinfo {author} {\bibfnamefont {J.}~\bibnamefont
  {Park}}\ and\ \bibinfo {author} {\bibfnamefont {B.}~\bibnamefont {Min}},\
  }\bibfield  {title} {\bibinfo {title} {Spatiotemporal plane wave expansion
  method for arbitrary space--time periodic photonic media},\ }\href
  {https://doi.org/10.1364/OL.411622} {\bibfield  {journal} {\bibinfo
  {journal} {Optics Letters}\ ,\ \bibinfo {pages} {484}} (\bibinfo {year}
  {2021})}\BibitemShut {NoStop}%
\bibitem [{\citenamefont {Sharabi}\ \emph {et~al.}(2021)\citenamefont
  {Sharabi}, \citenamefont {Lustig},\ and\ \citenamefont
  {Segev}}]{sharabi2021disordered}%
  \BibitemOpen
  \bibfield  {author} {\bibinfo {author} {\bibfnamefont {Y.}~\bibnamefont
  {Sharabi}}, \bibinfo {author} {\bibfnamefont {E.}~\bibnamefont {Lustig}},\
  and\ \bibinfo {author} {\bibfnamefont {M.}~\bibnamefont {Segev}},\ }\bibfield
   {title} {\bibinfo {title} {Disordered photonic time crystals},\ }\href@noop
  {} {\bibfield  {journal} {\bibinfo  {journal} {Physical Review Letters}\
  }\textbf {\bibinfo {volume} {126}},\ \bibinfo {pages} {163902} (\bibinfo
  {year} {2021})}\BibitemShut {NoStop}%
\bibitem [{\citenamefont {Ptitcyn}\ \emph {et~al.}(2019)\citenamefont
  {Ptitcyn}, \citenamefont {Mirmoosa},\ and\ \citenamefont
  {Tretyakov}}]{GregAtom}%
  \BibitemOpen
  \bibfield  {author} {\bibinfo {author} {\bibfnamefont {G.}~\bibnamefont
  {Ptitcyn}}, \bibinfo {author} {\bibfnamefont {M.}~\bibnamefont {Mirmoosa}},\
  and\ \bibinfo {author} {\bibfnamefont {S.}~\bibnamefont {Tretyakov}},\
  }\bibfield  {title} {\bibinfo {title} {Time-modulated meta-atoms},\
  }\href@noop {} {\bibfield  {journal} {\bibinfo  {journal} {Physical Review
  Research}\ }\textbf {\bibinfo {volume} {1}},\ \bibinfo {pages} {023014}
  (\bibinfo {year} {2019})}\BibitemShut {NoStop}%
\bibitem [{\citenamefont {Boltasseva}\ \emph {et~al.}(2024)\citenamefont
  {Boltasseva}, \citenamefont {Shalaev},\ and\ \citenamefont
  {Segev}}]{boltasseva2024photonic}%
  \BibitemOpen
  \bibfield  {author} {\bibinfo {author} {\bibfnamefont {A.}~\bibnamefont
  {Boltasseva}}, \bibinfo {author} {\bibfnamefont {V.}~\bibnamefont
  {Shalaev}},\ and\ \bibinfo {author} {\bibfnamefont {M.}~\bibnamefont
  {Segev}},\ }\bibfield  {title} {\bibinfo {title} {Photonic time crystals:
  from fundamental insights to novel applications: opinion},\ }\href@noop {}
  {\bibfield  {journal} {\bibinfo  {journal} {Optical Materials Express}\
  }\textbf {\bibinfo {volume} {14}},\ \bibinfo {pages} {592} (\bibinfo {year}
  {2024})}\BibitemShut {NoStop}%
\bibitem [{\citenamefont {Pacheco-Pe{\~n}a}\ \emph {et~al.}(2023)\citenamefont
  {Pacheco-Pe{\~n}a}, \citenamefont {Kiasat}, \citenamefont {Sol{\'\i}s},
  \citenamefont {Edwards},\ and\ \citenamefont {Engheta}}]{pacheco2023holding}%
  \BibitemOpen
  \bibfield  {author} {\bibinfo {author} {\bibfnamefont {V.}~\bibnamefont
  {Pacheco-Pe{\~n}a}}, \bibinfo {author} {\bibfnamefont {Y.}~\bibnamefont
  {Kiasat}}, \bibinfo {author} {\bibfnamefont {D.~M.}\ \bibnamefont
  {Sol{\'\i}s}}, \bibinfo {author} {\bibfnamefont {B.}~\bibnamefont
  {Edwards}},\ and\ \bibinfo {author} {\bibfnamefont {N.}~\bibnamefont
  {Engheta}},\ }\bibfield  {title} {\bibinfo {title} {Holding and amplifying
  electromagnetic waves with temporal non-foster metastructures},\ }\href@noop
  {} {\bibfield  {journal} {\bibinfo  {journal} {arXiv preprint
  arXiv:2304.03861}\ } (\bibinfo {year} {2023})}\BibitemShut {NoStop}%
\bibitem [{\citenamefont {Akbarzadeh}\ \emph {et~al.}(2018)\citenamefont
  {Akbarzadeh}, \citenamefont {Chamanara},\ and\ \citenamefont
  {Caloz}}]{akbarzadeh2018inverse}%
  \BibitemOpen
  \bibfield  {author} {\bibinfo {author} {\bibfnamefont {A.}~\bibnamefont
  {Akbarzadeh}}, \bibinfo {author} {\bibfnamefont {N.}~\bibnamefont
  {Chamanara}},\ and\ \bibinfo {author} {\bibfnamefont {C.}~\bibnamefont
  {Caloz}},\ }\bibfield  {title} {\bibinfo {title} {Inverse prism based on
  temporal discontinuity and spatial dispersion},\ }\href@noop {} {\bibfield
  {journal} {\bibinfo  {journal} {Optics letters}\ }\textbf {\bibinfo {volume}
  {43}},\ \bibinfo {pages} {3297} (\bibinfo {year} {2018})}\BibitemShut
  {NoStop}%
\end{thebibliography}%
\end{document}